\documentclass{article}
\usepackage{graphicx}

\begin{document}

\title{ Optical-parametric-oscillator solitons driven by the third harmonic}
\author{Vitaly Lutsky and Boris A. Malomed \\
Department of Interdisciplinary Studies\\
Faculty of Engineering, Tel Aviv University\\
Tel Aviv 69978, Israel}
\maketitle

\begin{center}
\textbf{Abstract}
\end{center}

We introduce a model of a lossy second-harmonic-generating ($\chi
^{(2)}$) cavity externally pumped at the third harmonic, which
gives rise to driving terms of a new type, corresponding to a
\textit{cross-parametric} gain. The equation for the
fundamental-frequency (FF) wave may also contain a quadratic
self-driving term, which is generated by the cubic nonlinearity of
the medium. Unlike previously studied phase-matched models of
$\chi ^{(2)}$ cavities driven at the second harmonic (SH) or at
FF, the present one admits an exact analytical solution for the
soliton, at a special value of the gain parameter. Two families of
solitons are found in a numerical form, and their stability area
is identified through numerical computation of the perturbation
eigenvalues (stability of the zero solution, which is a necessary
condition for the soliton's stability, is investigated in an
analytical form). One family is a continuation of the special
analytical solution. At given values of parameters, one soliton is
stable and the other one is not; they swap their stability at a
critical value of the mismatch parameter. The stability of the
solitons is also verified in direct simulations, which demonstrate
that the unstable pulse rearranges itself into the stable one, or
into a delocalized state, or decays to zero. A soliton which was
given an initial boost $C$ starts to move but quickly comes to a
halt, if the boost is smaller than a critical value
$C_{\mathrm{cr}}$. If $C>C_{\mathrm{cr}}$, the boost destroys the
soliton (sometimes, through splitting into two secondary pulses).
Interactions between initially separated solitons are investigated
too. It is concluded that stable solitons always merge into a
single one. In the system with weak loss, it appears in a
vibrating form, slowly relaxing to the static shape. With stronger
loss, the final soliton emerges in the stationary form.

\begin{center}
\end{center}

\newpage

\section{Introduction}

In the vast family of optical solitons, an important niche is
occupied by solitary waves in cavities, supported by the quadratic
($\chi ^{(2)}$) nonlinearity of the degenerate
optical-parametric-oscillator (OPO) type
\cite{ReviewItaly,WeissReview}. The intrinsic loss in the cavity
should be compensated by an external pump field $E$, which in most
cases is supplied at the second harmonic (SH)
\cite{Lugiato:downconversion}. Such an arrangement is frequently
referred to as \textit{downconversion}, and is described by the
model in which either the evolution equation for the SH field $v$
explicitly contains a constant driving term $\sim E$,
\cite{Trillo,Falk:down}, or, equivalently, the equation for the
fundamental-frequency (FF) wave $u$ includes a parametric-gain
term $\sim Eu^{\ast}$, where the asterisk stands for the complex
conjugation \cite{Longhi,Staliunas,Skryabin}. Alternatively, the
cavity may be externally pumped at the FF, which is referred to as
\textit{upconversion}, the respective model including a constant
driving term in the FF equation \cite{Falk:up}. For both cases,
families of one-dimensional solitons and their stability have been
investigated in detail, see Refs. \cite{Trillo}-\cite{Alan} and
references therein. The stability of cavity solitons was tested in
a direct experiment using the photorefractive nonlinearity
\cite{stability-experiment}.

It should be mentioned that, as the cavity models are dissipative ones, the
solitary pulses found in these models are not solitons in the rigorous
sense. Nevertheless, this term is broadly applied to them, therefore we use
it in this paper too.

The analysis of solitons in OPO models was extended in various directions,
including the study of moving solitons \cite{moving}, interactions between
them \cite{interactions}, nondegenerate second-harmonic-generation (SHG)
which involves two FF waves with orthogonal polarizations \cite{type-II},
general three-wave interactions \cite{3-wave-solitons}, the use of the
quasi-phase-matching technique \cite{QPM}, two-dimensional solitons (see,
e.g., Ref. \cite{2D}), etc. Besides their significance as the subject of
fundamental research, OPO cavity solitons also have a potential for the
design of rewritable multi-pixel optical-memory patterns \cite{multi-pixel}.

In this work, we aim to propose and analyze another possibility to
drive dissipative cavities with the SHG nonlinearity, namely,
through a phase-matched third-harmonic (TH) pump wave, $w$.
Obviously, the corresponding driving terms in the equations for
the FF and SH fields $u$ and $v$, which are generated by the
parametric interaction of these fields with the TH pump, are
proportional, respectively, to $wv^{\ast }$ and $wu^{\ast }$.
Thus, this model includes a new feature, the
\textit{cross-parametric gain}, in the system of coupled FF and SH
waves, and a problem of straightforward interest is to study
solitons that can be supported by this type of the gain in the
lossy SHG setting, especially as concerns stability of the
solitons. Besides that, we will also take into regard a
possibility of an additional quadratically nonlinear parametric
self-driving term in the equation for the FF field $u$, in the
form of $w\left( u^{\ast }\right) ^{2}$, which may be induced by
the same TH pump through the $\chi ^{(3)}$ (cubic) nonlinearity.
In this work, we consider only bright solitons; dark solitons, and
patterns in the form of domain walls (see, e.g., Ref.
\cite{LeBerre}) may also be possible in the model including the TH
drive.

As concerns the mutual phase matching between the FF, SH, and TH
fields, which is implied in the model, it was demonstrated, in
another context, that matching of this type may take place in the
so-called multi-step $\chi ^{(2)} $ systems \cite{multistep}.
However, the subject of Ref. \cite{multistep} was the
corresponding three-wave solitons, rather than pumping of the FF
and SH fields through the TH wave.

The paper is organized as follows. In Section II, we give the
formulation of the model, and find particular exact solutions for
the solitons in an analytical form. In the same section, we also
investigate stability conditions for the zero solution, which is a
necessary prerequisite for the stability of solitons. In Section
III, we find two families of general soliton solutions in a
numerical form. One family is a direct continuation of the exact
solution, while the other one is different. As well as the
analytically found soliton, they always have a single-humped
shape, but, unlike the constant-phase exact solutions, they
feature intrinsic chirp. Stability regions for the solitons are
identified, in the system's parameter space, through computation
of the corresponding eigenvalues for small perturbations. One of
the solitons is stable, and the other one is not; they swap the
stability via a bifurcation at a critical value of the mismatch
parameter. The shape of the stability regions is quite nontrivial;
in the model with weak losses, the stable solitons may have
complex eigenvalues, corresponding to weakly damped intrinsic
oscillatory modes. The stability is also verified in direct
simulations, with a conclusion that the unstable soliton
rearranges into the stable one, or into a delocalized spatially
periodic state, or, sometimes, it decay to zero. Also in Section
III, we show that attempts to produce moving solitons fail: if the
soliton is initially boosted by lending it a speed $C$, it quickly
stops, provided that $C$ is smaller than a critical value
$C_{\mathrm{cr}}$; the boost with $C>$ $C_{\mathrm{cr}}$ destroys
the soliton. Section IV deals with interactions between two stable
solitons, initially placed at some distance. They merge into a
single soliton, which emerges with slowly fading intrinsic
vibrations in the weakly dissipative model, or immediately in the
stationary form if the loss parameter is larger. Section V
discusses possible extensions of the work and concludes the paper.

\section{The model and exact results}

\label{sec:The_model}

Equations which describe the degenerate $\chi ^{(2)}$ interaction between
the FF and SH fields $u(x,t)$ and $v(x,t)$ in a one-dimensional lossy cavity
in the presence of an additional pump TH wave $w_{0}$, whose depletion is
negligible, are straightforward to derive:
\begin{eqnarray}
i\omega _{0}u_{t}+\frac{1}{2}u_{xx}+\chi ^{(2)}u^{\ast }v &=&\left(
q_{1}-i\alpha _{1}\right) u+iw_{0}v^{\ast }+i\chi ^{(3)}w_{0}\left( u^{\ast
}\right) ^{2},  \label{FF} \\
2i\omega _{0}v_{t}+\frac{1}{2}v_{xx}+\frac{1}{2}\chi ^{(2)}u^{2} &=&2\left(
q_{2}-i\alpha _{2}\right) v+iw_{0}u^{\ast }.  \label{SH}
\end{eqnarray}Here, $\omega _{0}$ is the fundamental carrier frequency, the coefficient
amenable for the three-wave coupling between the FF, SH and TH
fields is absorbed into $w_{0}$, $q_{1}$ and $q_{2}$ are real
detuning coefficients at the FF and SH, and $\alpha _{1}$ and
$\alpha _{2}$ are the loss coefficients for the same fields. In
this paper, we focus on the most natural case, $\alpha _{1}=\alpha
_{2}$, but the analysis has demonstrated that the results do nor
differ in any noticeable aspect for $\alpha _{1}\neq \alpha _{2}$.
The last term in Eq. (\ref{FF}) takes into regard the
above-mentioned possibility of the nonlinear parametric
self-driving of the FF field, which can directly couple to the TH
pump wave in the presence of the cubic nonlinearity. The factors
of $i$ in front of the cross-parametric-driving terms in Eqs.
\ref{FF}) and (\ref{SH}), while $w_{0}$ is assumed real, can
always be fixed, defining phase shifts between the corresponding
waves.

After obvious normalizations (in particular, the SH field is
rescaled with $u\rightarrow \sqrt{2}u$, so as to keep the
coefficients in front of the cross-driving terms equal in the two
equations), Eqs. (\ref{FF}) and (\ref{SH}) can be cast in the
following form:
\begin{eqnarray}
iu_{t}+\frac{1}{2}u_{xx}+u^{\ast }v &=&\left( 1-i\alpha _{1}\right)
u+i\alpha _{0}v^{\ast }+i\beta \left( u^{\ast }\right) ^{2},  \label{u} \\
iv_{t}+\frac{1}{4}v_{xx}+\frac{1}{2}u^{2} &=&\left( 2q-i\alpha _{1}\right)
v+i\alpha _{0}u^{\ast },  \label{v}
\end{eqnarray}where $\alpha _{0}$ and $\beta $ are the effective pumping coefficients,
both proportional to $w_{0}$, the mismatch coefficient in Eq.
(\ref{u}) is normalized to be $1$ (a different variant of the
model is obtained by fixing the latter coefficient to be $-1$, but
the existence of bright solitons is not expected in that case).
The notation for the fields $u$ and $v$ and the coefficient
$\alpha _{1}$ was not altered, although they were rescaled (recall
we assume $\alpha _{2}=\alpha _{1}$). By means of a phase shift of
$u $ and $v$, $\alpha _{0}$ may always be made real, which we
assume below, while $\beta $ is, generally speaking, a complex
coefficient.

Obviously, Eqs. (\ref{u}) and (\ref{v}) cannot give rise to stable localized
solutions unless the zero solution, $u=v=0$, is stable. Linearizing the
equations, an elementary calculation yields the following stability
condition for the zero solution:
\begin{eqnarray}
\alpha _{0}^{2} &\leq &\alpha _{1}^{2}+\left( 1+2q\right) ^{2}/4,~\mathrm{if}~~1+2q>0;  \nonumber \\
\alpha _{0}^{2} &\leq &\alpha _{1}^{2},~\mathrm{if}~~1+2q<0.
\label{zero-stability}
\end{eqnarray}The consideration of the linearized version of Eqs. (\ref{u}) and (\ref{v})
makes it also possible to predict the asymptotic form of the exponentially
decaying tails of the soliton solution at $|x|~\rightarrow \infty $,
\begin{equation}
\left( u,v^{\ast }\right) \sim \exp \left( -\lambda |x|\right) ,  \label{exp}
\end{equation}where a (generally) complex constant $\lambda $ (it is defined so that its
real part is positive) is to be found from the equation
\begin{equation}
\left( \frac{1}{2}\lambda ^{2}-1+i\alpha _{1}\right) \left(
\frac{1}{4}\lambda ^{2}-2q-i\alpha _{1}\right) =\alpha _{0}^{2}~.
\label{lambda}
\end{equation}The complex structure of $\lambda $ implies that the soliton must have a
nontrivial intrinsic phase structure, which will be studied in detail below.

Particular exact soliton solutions to Eqs. (\ref{u}) and (\ref{v}) can be
sought for as
\begin{equation}
u_{0}(x)=A\,e^{i\phi }\mathrm{sech}^{2}\left( \kappa x\right)
,\,\,v_{0}(x)=B~e^{2i\phi }\mathrm{sech}^{2}\left( \kappa x\right) ,
\label{soliton}
\end{equation}where $\kappa $, $\phi $ and $A$, $B$ are real constants. This ansatz may
produce solutions in the case of $\beta =0$ [no self-driving
quadratic term in Eq. (\ref{u})]. Then, the soliton's parameters
are obtained by direct substitution of the expressions
(\ref{soliton}) into Eqs. (\ref{u}) and (\ref{v}):\begin{equation}
\kappa ^{2}=1-2q,  \label{simplekappa}
\end{equation}\begin{equation}
A=B=3\kappa ^{2},  \label{BA^2}
\end{equation}\begin{equation}
\cos \left( 3\phi \right) =\alpha _{1}/\alpha _{0}\,,~\sin \left( 3\phi
\right) =\left( 1-4q\right) /\alpha _{0}\,~.  \label{cos}
\end{equation}Due to the identity $\sin ^{2}\left( 3\phi \right) +\cos ^{2}\left( 3\phi
\right) \equiv 1$, Eqs. (\ref{cos}) give rise to an additional constraint on
the parameters of the model which is necessary for the existence of the
exact solution in the above form,
\begin{equation}
\alpha _{0}=\left( \alpha _{0}\right) _{\mathrm{exact}}\equiv \sqrt{\alpha
_{1}^{2}+\left( 1-4q\right) ^{2}}.  \label{alpha0}
\end{equation}

Equation (\ref{alpha0}) determines the value of the cross-parametric-gain
which is necessary to support the exact soliton solution. Thus, the two
conditions, $\beta =0$ and (\ref{alpha0}), must be imposed on the parameters
of Eqs. (\ref{u}) and (\ref{v}) to provide for the existence of the
analytical solution for the soliton in the simple form (\ref{soliton}) (in
fact, the condition $\alpha _{1}=\alpha _{2}$, which was adopted above, is
also necessary for the existence of the exact soliton in the present form).

It is relevant to notice that, comparing Eq. (\ref{soliton}) with the
asymptotic waveform (\ref{exp}), one can identify, for the present solution,
$\lambda =2\kappa $. Then, taking into regard Eq. (\ref{alpha0}), it is easy
to verify that this $\lambda $ indeed satisfies Eq. (\ref{lambda}).

A necessary stability criterion for the exact soliton solution can be
obtained, inserting the relation (\ref{alpha0}) into the stability condition
(\ref{zero-stability}) for the zero solution. After a simple algebra, it
takes the form
\begin{equation}
1/10\leq q\leq 1/2~.  \label{interval}
\end{equation}

\begin{figure}[tbph]
\begin{center}
$\begin{array}{c@{\hspace{0.5in}}c}
\includegraphics[width=2.5in]{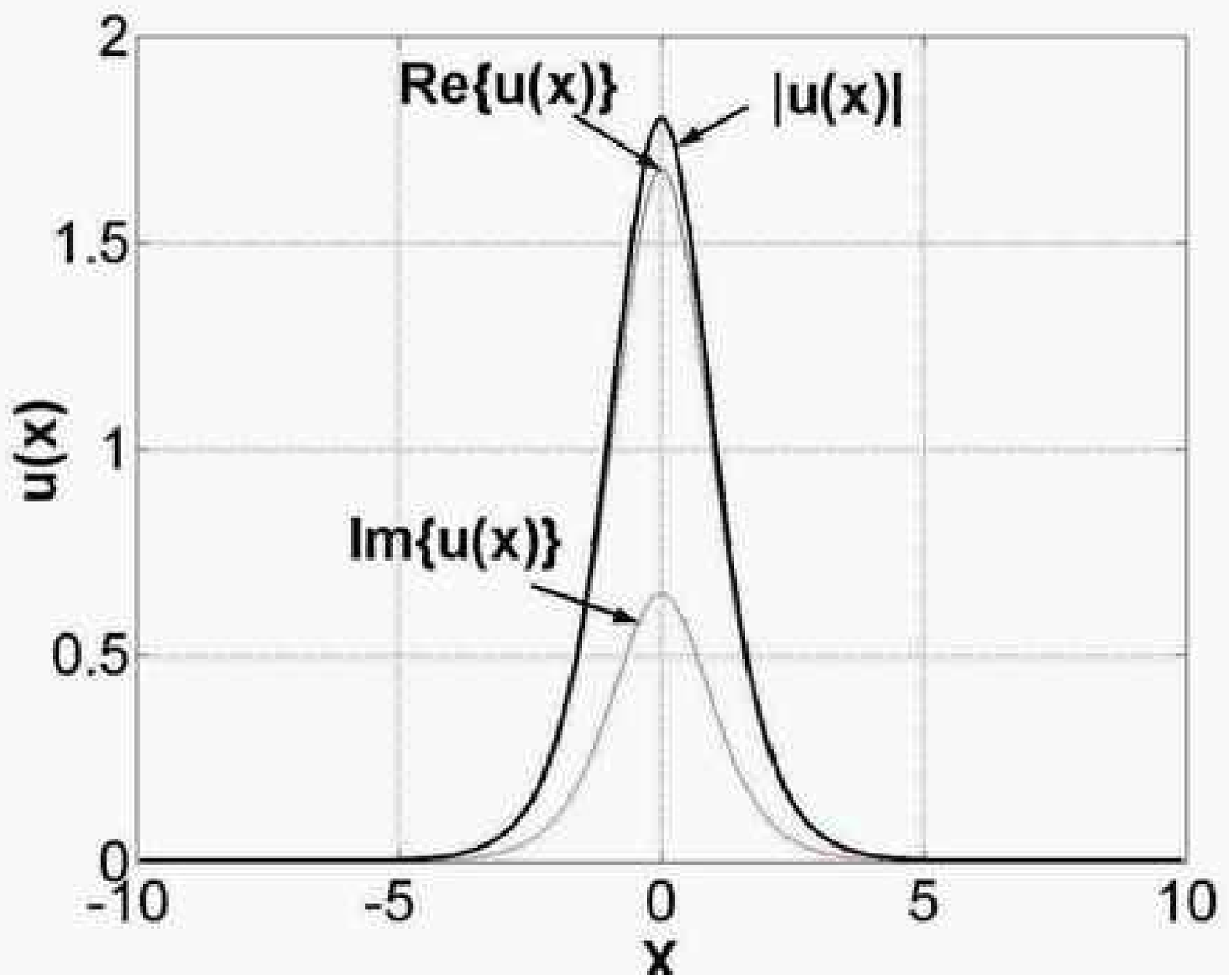} & \includegraphics[width=2.5in]{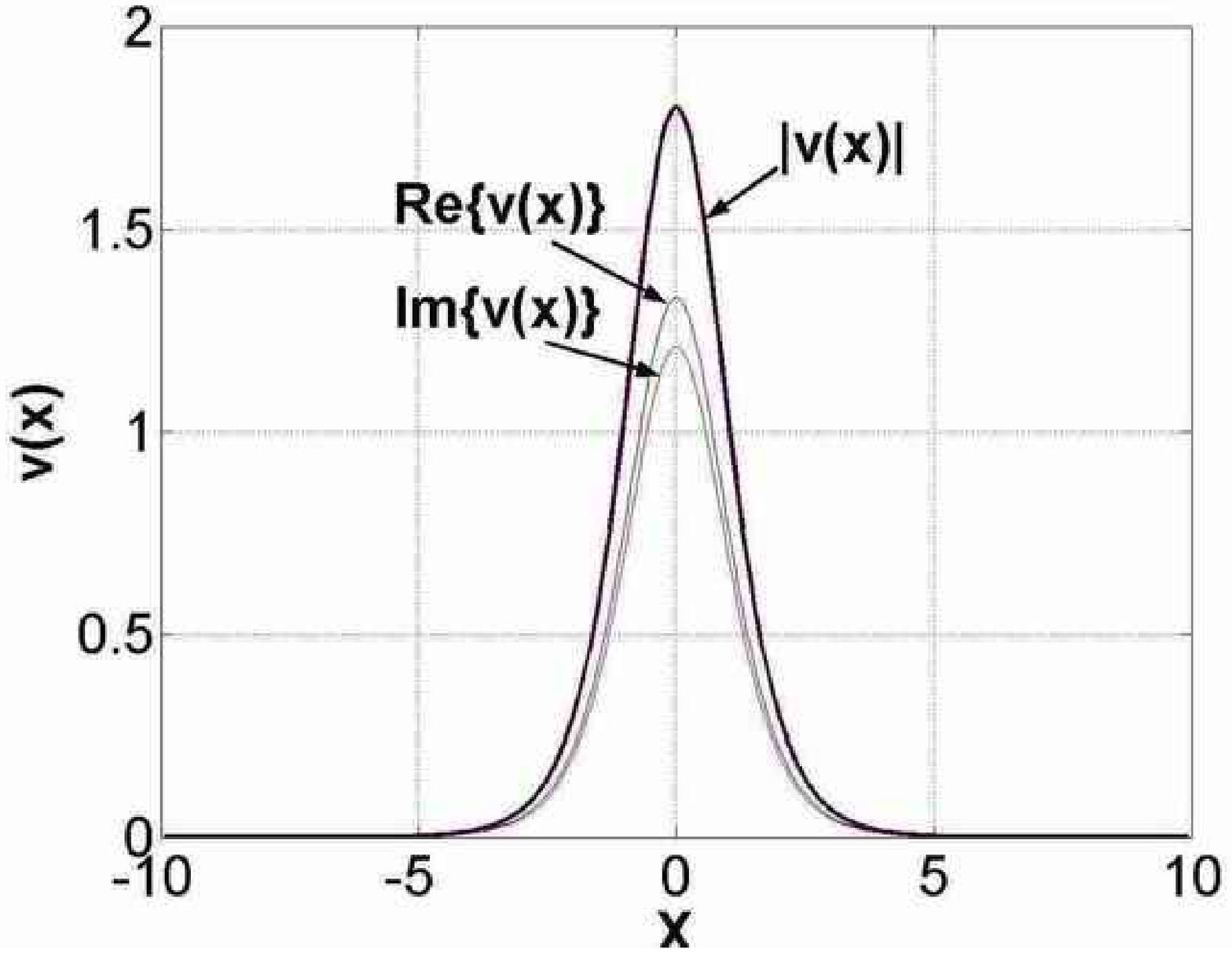} \\[0.4cm]
\mbox{\bf (a)} & \mbox{\bf (b)}\end{array}$\end{center} \caption{A
typical shape of the soliton's FF (a) and SH components (b), as
given by the exact analytical expressions (\protect\ref{soliton}),
(\protect \ref{BA^2}), (\protect\ref{simplekappa}), and
(\protect\ref{cos}) for $\protect\beta =0$, $\protect\alpha
_{1}=0.1$, and $q=0.2$; accordingly, Eq. (\protect\ref{alpha0})
yields, in this case, $\left( \protect\alpha _{0}\right)
_{\mathrm{exact}}=0.1\protect\sqrt{5}\approx 0.224$.}
\label{Fig:exact}
\end{figure}

A typical example of the exact soliton given by the above
expressions is shown in Fig. \ref{Fig:exact}. Below, numerical
results will be produced for $\alpha _{0}\neq \left( \alpha
_{0}\right) _{\mathrm{exact}}$ and $\beta \neq 0$, and they will
be compared to the shape shown in Fig. \ref{Fig:exact}.

It is relevant to mention that the previously considered
phase-matched OPO models with the parametric gain of the
downconversion type (provided by the SH pump) did not produced
exact solutions similar to the present one. Exact solutions were
only obtained in the case of large detuning
\cite{Longhi,exact-solution-cascading-limit}. In that limit, the
SH field can be eliminated, and the remaining FF equation amounts
to a parametrically driven damped cubic nonlinear Schr\"{o}dinger
equation, which has a pair of well-known exact solitary-pulse
solutions (see, e.g., Ref. \cite{Barash}). On the other hand,
nongeneric exact analytical solutions for solitons (which
explicitly contain the chirp) were found in a model of a two-core
$\chi ^{(2)}$ system of the Ginzburg-Landau type, in which
intrinsic gain was set in the nonlinear core, and a linearly
coupled additional lossy core played a stabilizing role
\cite{Lucian}. However, the gain in that system was not of the
parametric type.

\section{\textbf{The soliton family: numerical results}}

The analytical solutions were found above in the special case only, and even
in that case, full stability analysis requires the use of numerical methods.
In this section, we aim to construct a general family of soliton solutions
in a numerical form, and then study their stability. A possibility of the
existence of moving solitons will be considered too.

\subsection{\textbf{Stationary solitons}}

\begin{figure}[tbph]
\begin{center}
$\begin{array}{c@{\hspace{0.5in}}c}
\includegraphics[width=2.5in]{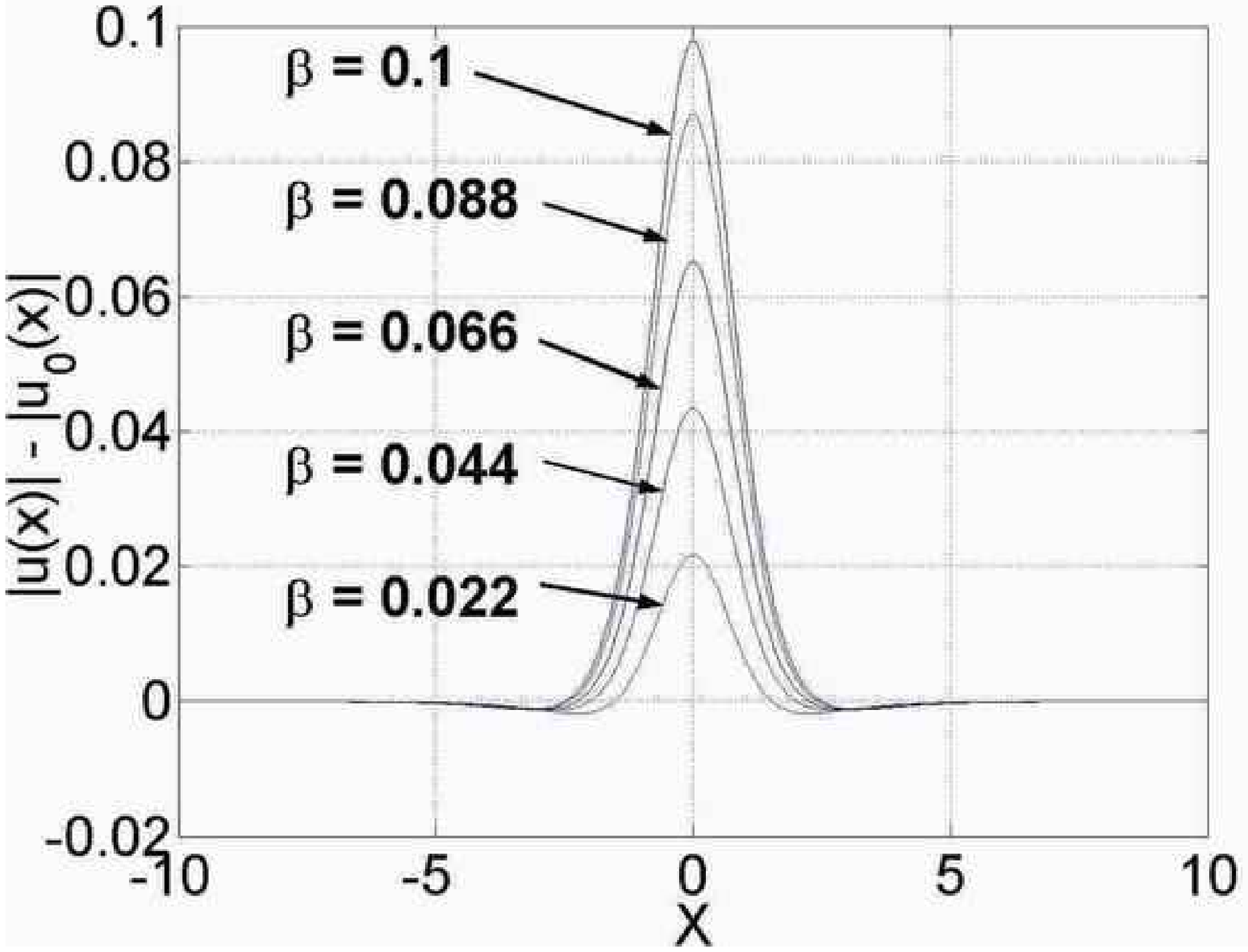} & \includegraphics[width=2.5in]{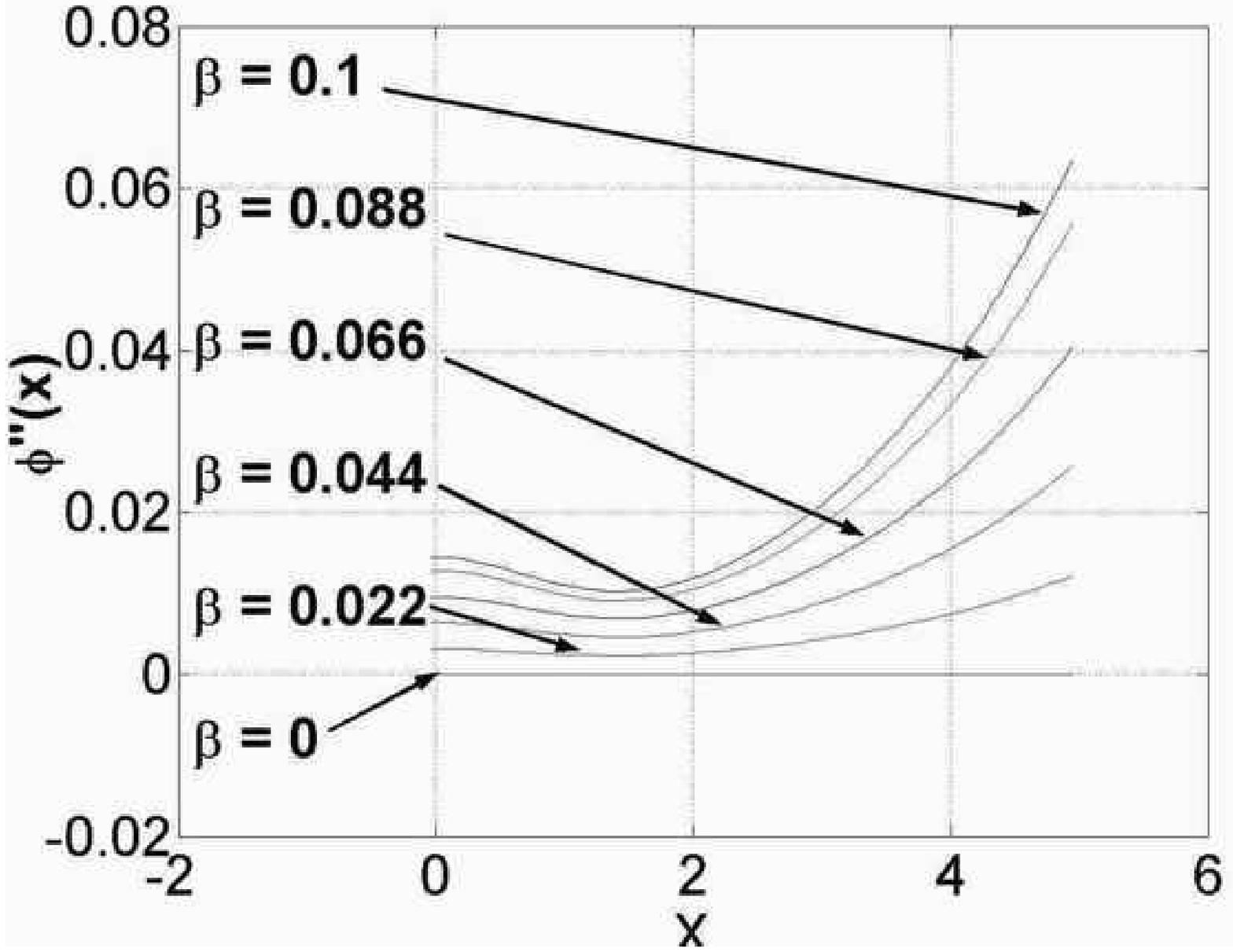} \\[0.4cm]
\mbox{\bf (a)} & \mbox{\bf (b)} \\
\includegraphics[width=2.5in]{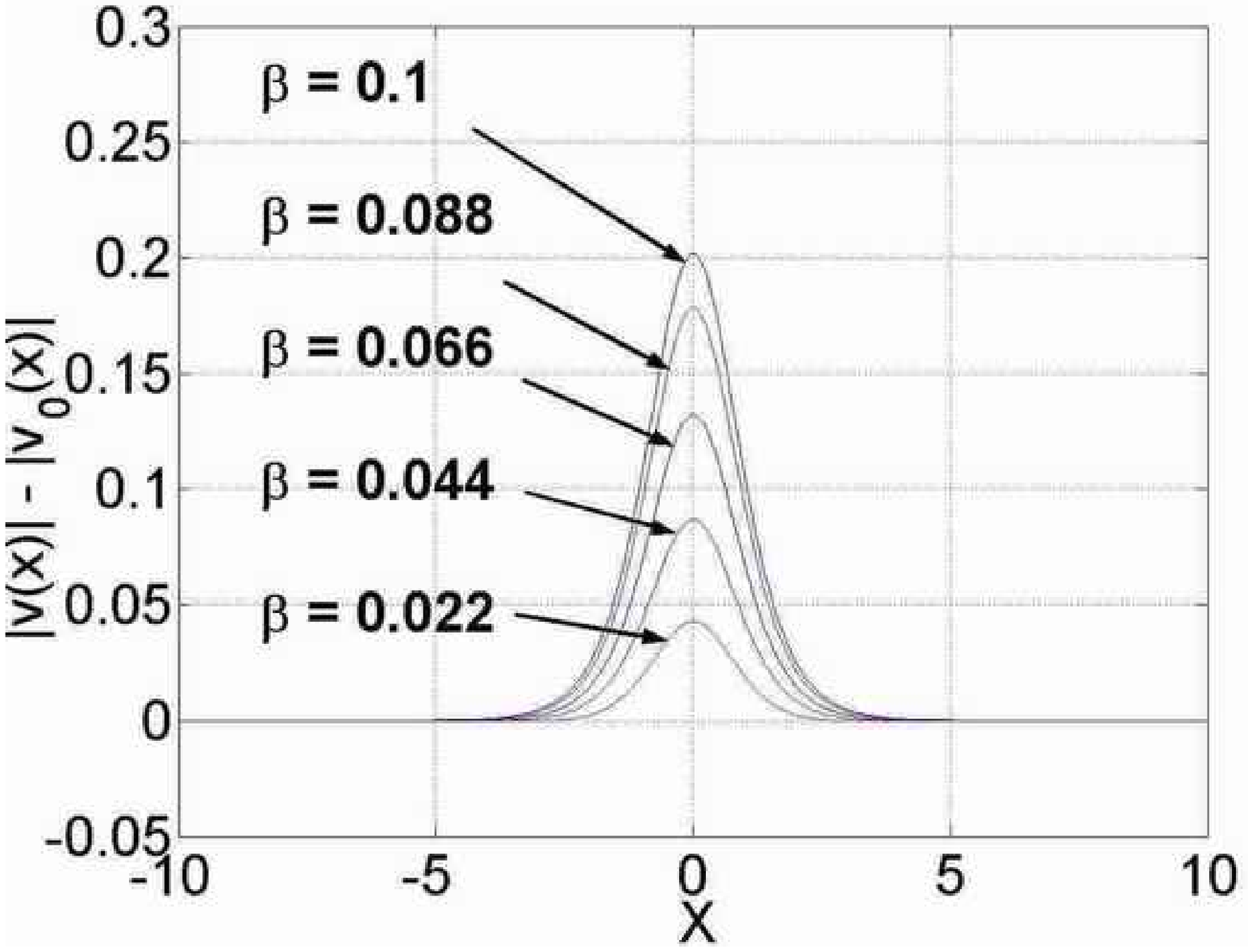} & \includegraphics[width=2.5in]{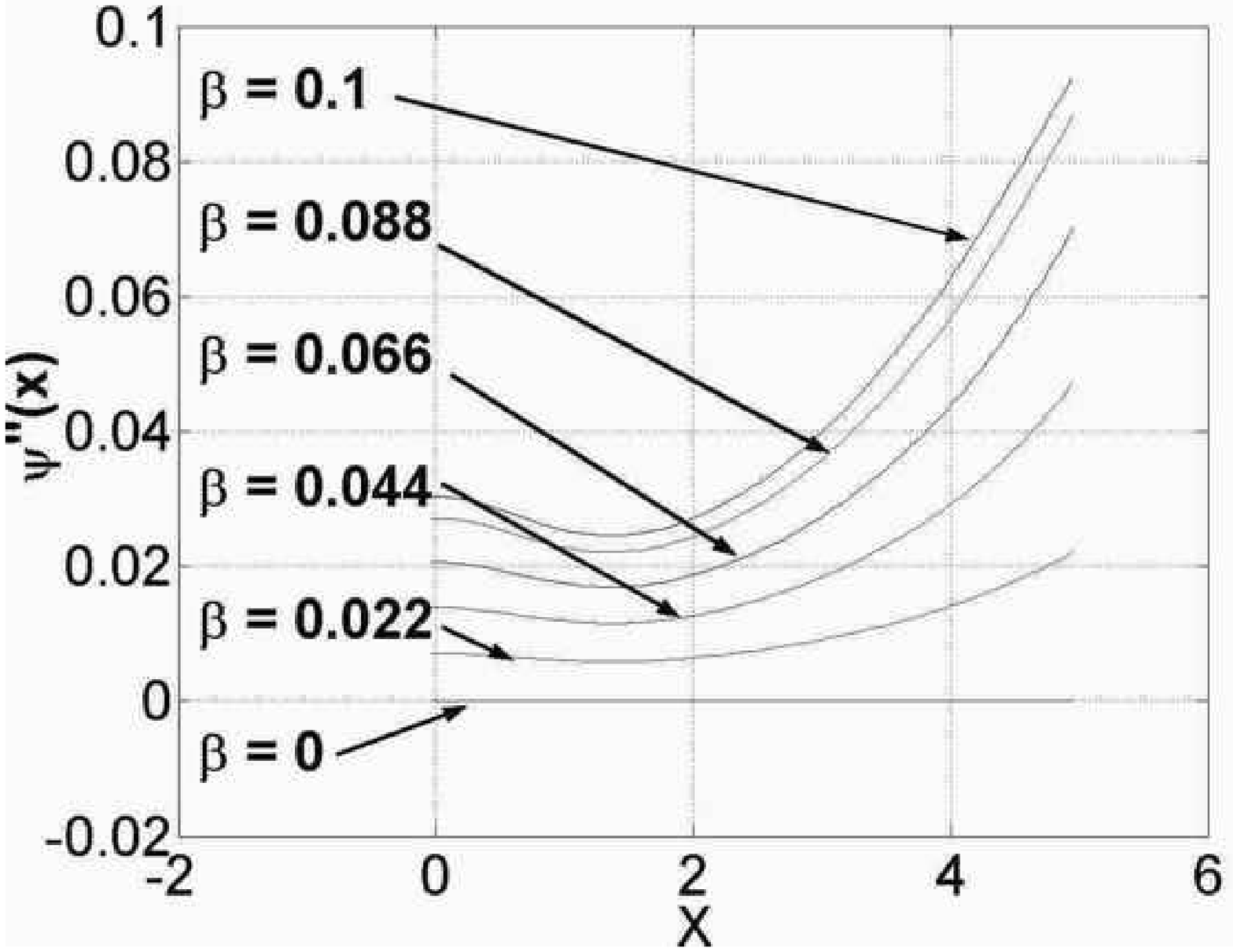} \\[0.4cm]
\mbox{\bf (c)} & \mbox{\bf (d)}\end{array}$\end{center}
\caption{The differences, $|u(x)|-|u_{0}(x)|$ and
$|v(x)|-|v_{0}(x)|$, between the amplitude distributions in the FF
(a) and SH (b) components of the numerically found generic
solitons and the exact analytical solution. The panels (b) and (d)
display the phase chirp, $\protect\phi ^{\prime \prime }$ and
$\protect\psi ^{\prime \prime }$, in the FF and SH components of
the generic soliton solutions (the chirp distributions are shown
on the half-axis, in view of the soliton's symmetry, and only in
the region of $x$ where the soliton is actually located). The
fixed parameters are $q=0.2$, $\protect\alpha _{1}=0.1$, and, for
this figure, the corresponding value $\left( \protect\alpha
_{0}\right) _{\mathrm{exact}}=0.224$, as per Eq.
(\protect\ref{alpha0}), is chosen. Note that the full shapes,
$|u(x)|$ and $|v(x)|$, of all the solitons in both components are
strictly single-humped ones.} \label{Fig:generic0}
\end{figure}

\begin{figure}[tbph]
\begin{center}
$\begin{array}{c@{\hspace{0.5in}}c}
\includegraphics[width=2.5in]{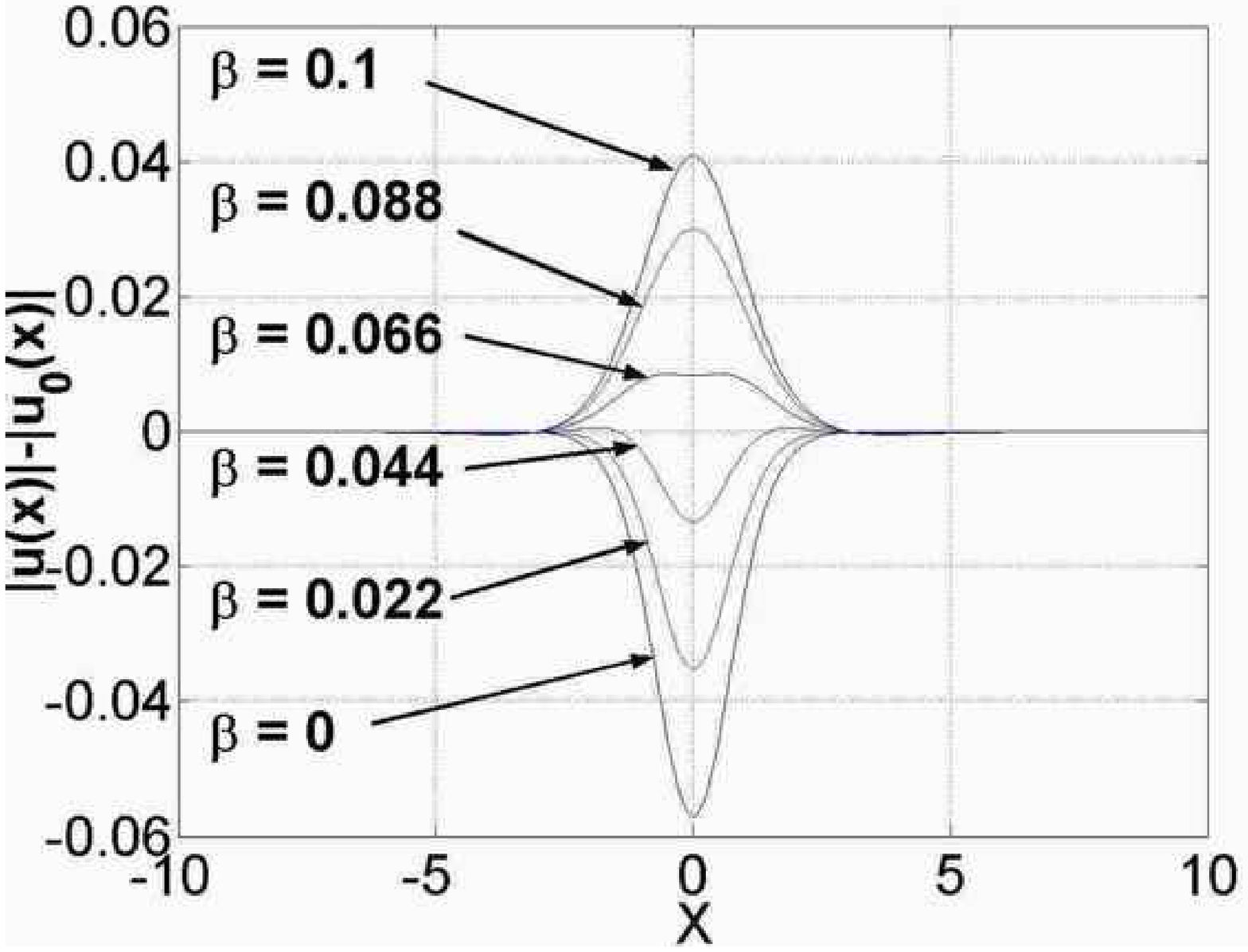} & \includegraphics[width=2.5in]{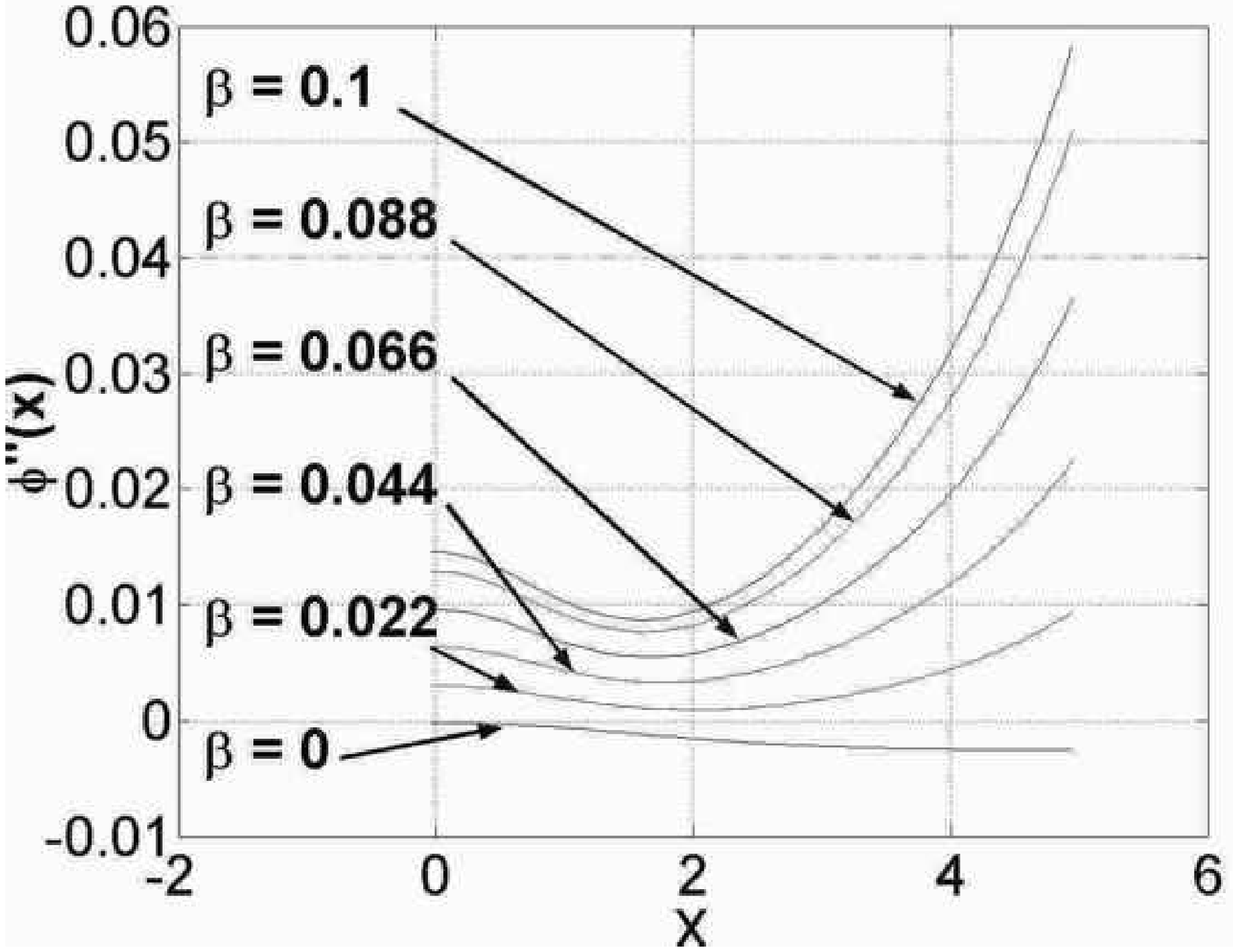} \\[0.4cm]
\mbox{\bf (a)} & \mbox{\bf (b)} \\
\includegraphics[width=2.5in]{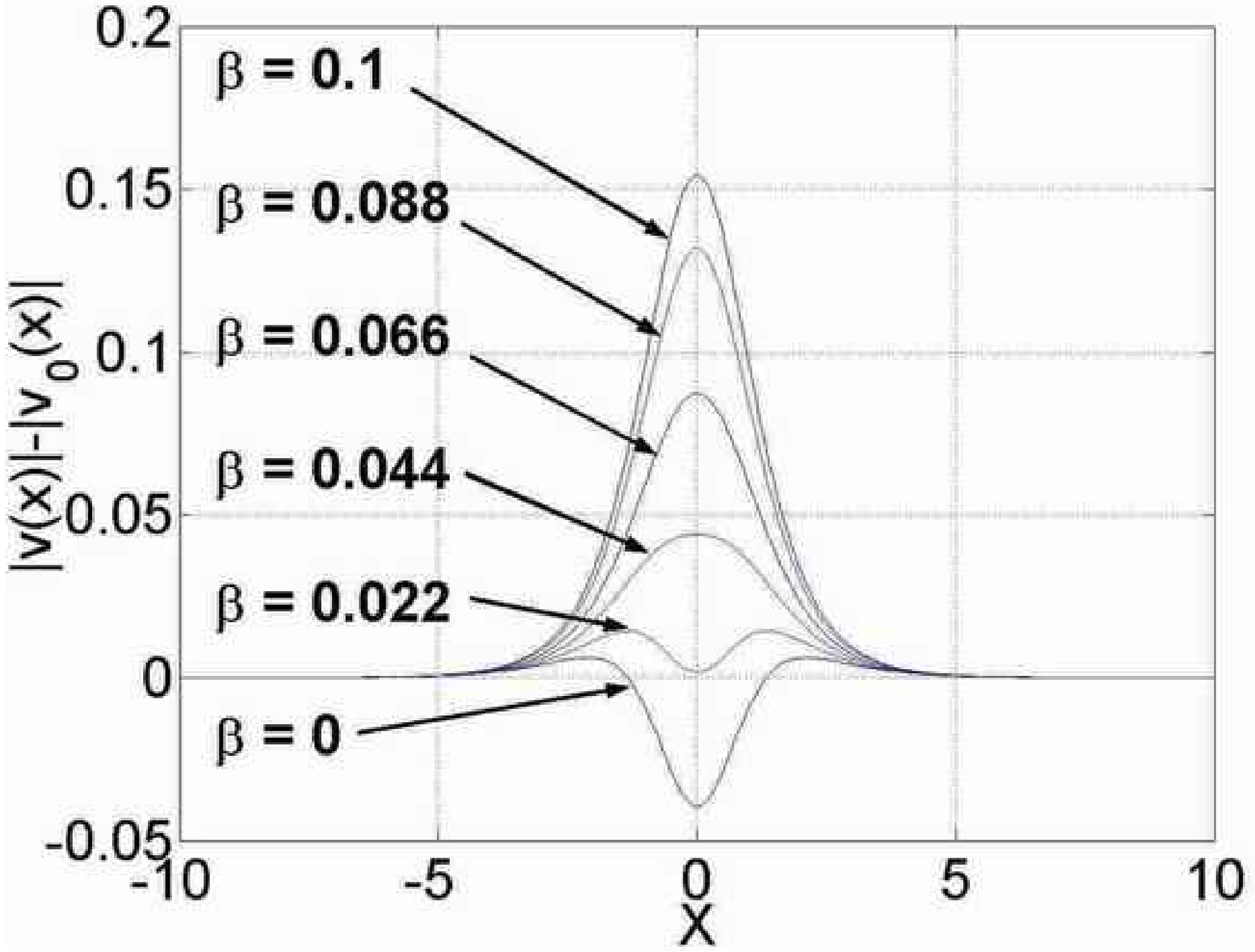} & \includegraphics[width=2.5in]{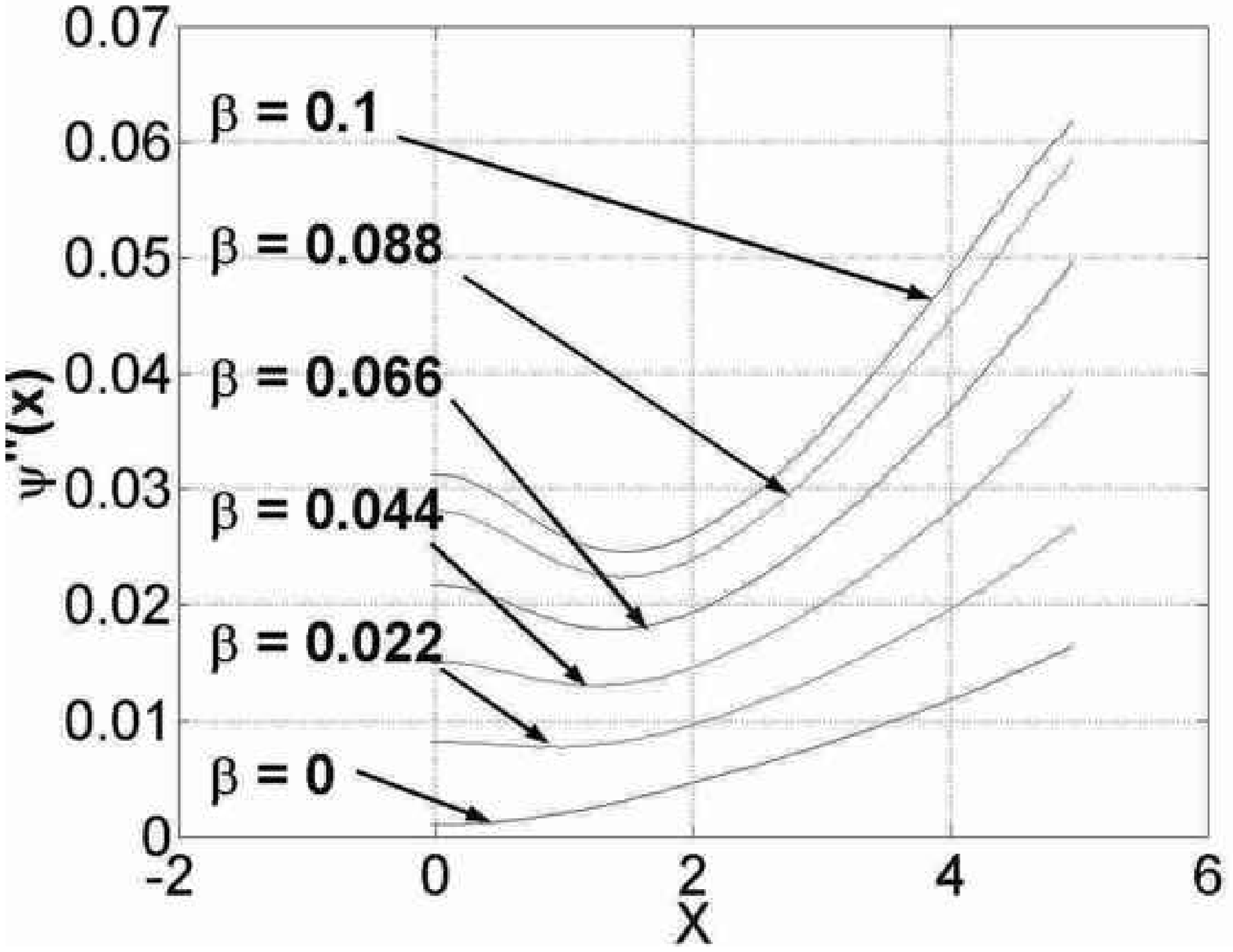} \\[0.4cm]
\mbox{\bf (c)} & \mbox{\bf (d)}\end{array}$\end{center}
\caption{The same as in Fig. \protect\ref{Fig:generic0}, but for
$\protect\alpha _{0}=0.9\left( \protect\alpha _{0}\right)
_{\mathrm{exact}}=0.201$.} \label{Fig:generic}
\end{figure}

A family of soliton solutions to the stationary version of Eqs.
(\ref{u}) and (\ref{v}), with $u_{t}=v_{t}=0$, was constructed by
means of a continuation procedure (based on the Newton's numerical
method), starting with the exact solutions in the form given by
Eqs. (\ref{soliton}) - (\ref{cos}), which are valid in the case of
$\beta =0$ and $\alpha _{0}=\left( \alpha _{0}\right)
_{\mathrm{exact}}$, and then gradually varying both $\alpha _{0}$
and $\beta $. As a result, it was concluded that the shape of the
soliton in both the FF and SH components, $|u(x)|$ and $|v(x)|$,
does not not vary much in comparison with that of the exact
solution, while a new feature is an intrinsic phase structure of
the soliton's wave field, characterized by nonzero
\textit{chirps}, $\phi ^{\prime \prime }$ and $\psi ^{\prime
\prime }$, where $\phi (x)$ and $\psi (x)$ are phases of the
fields $u(x)$ and $v(x)$. These features are illustrated by Figs.
\ref{Fig:generic0} and \ref{Fig:generic}, which display two
generic subfamilies of the numerically found stationary soliton
solutions, obtained by varying the self-driving coefficient $\beta
$ at fixed values of the cross-parametric-drive coefficient
$\alpha _{0}$ (in these figures, only real values of $\beta $ are
included; complex $\beta $, which do not produce effects
drastically different from those found for real $\beta $, will be
briefly considered below). Figure \ref{Fig:generic0} corresponds
to $\alpha _{0}=\left( \alpha _{0}\right) _{\mathrm{exact}}$,
which is the value that gives rise to the exact solution (for
$\beta =0$), see Eq. (\ref{alpha0}), and Fig. \ref{Fig:generic}
presents the situation at a different value of $\alpha _{0}$,
namely, $\alpha _{0}=0.9\left( \alpha _{0}\right)
_{\mathrm{exact}}$. In these figures, the amplitude profiles are
shown through their differences from those corresponding to the
exact soliton solution (\ref{soliton}) - (\ref{cos}), taken at
$\beta =0$ and $\alpha _{0}=\left( \alpha _{0}\right)
_{\mathrm{exact}}$, because full profiles are too close to each
other.

At other values of $\alpha _{0}$, the soliton solutions are quite similar to
those displayed in Figs. \ref{Fig:generic0} and \ref{Fig:generic}. We stress
that, while the amplitude-profile differences displayed in these figures may
have a relatively complex form, the full profiles of $|u(x)|$ and $|v(x)|$
always keep the simple single-humped shape.

\subsection{\textbf{Stability of the stationary solitons}}

\begin{figure}[tbph]
\begin{center}
\includegraphics[width=2.5in]{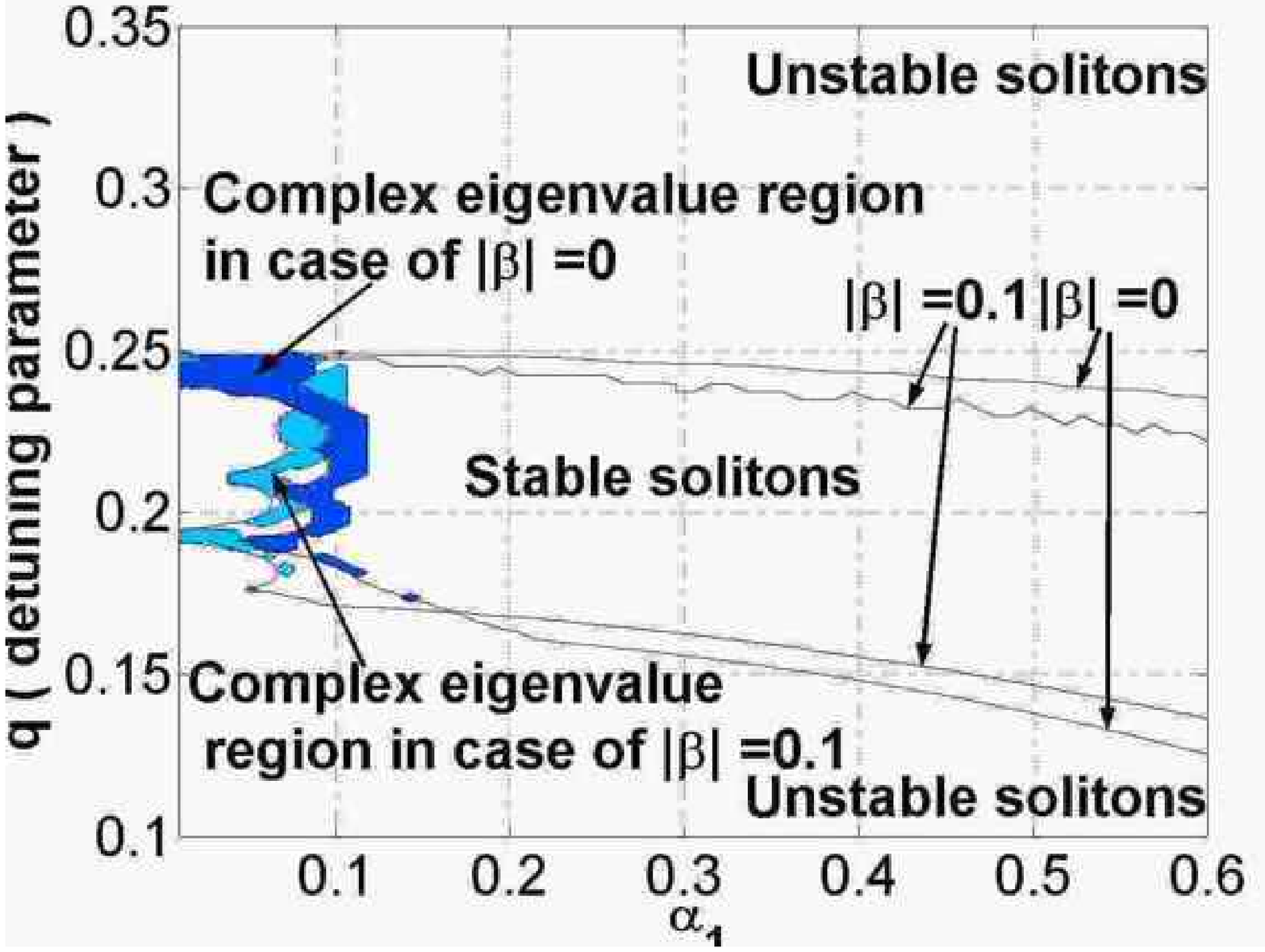}
\end{center}
\caption{The stability maps for the family of the exact soliton
solutions (\protect\ref{soliton}) - (\protect\ref{cos}),
corresponding to $\protect\beta =0$ and $\protect\alpha
_{0}=\left( \protect\alpha _{0}\right) _{\mathrm{exact}}$, see Eq.
(\protect\ref{alpha0}), and for the solitons found numerically for
the same $\protect\alpha _{0}$ but $\protect\beta =0.1$. The
stability is identified from the computation of the stability
eigenvalues (and verified by direct simulations). The shading,
bordered by the dashed lines, marks stability sub-regions in which
complex eigenvalues were found, while, in the unshaded areas, all
the eigenvalues are real.} \label{fig:Beta_only_graph}
\end{figure}

The most fundamental approach to the investigation of the
stability of stationary solitons is based on computation of
stability eigenvalues within the framework of the system of
linearized equations for infinitesimal perturbations. We have
performed this computation through the corresponding Jacobian
matrix of the linearized system. The results can be conveniently
summarized in the form of maps showing stable and unstable regions
in the system's parameter planes.\textbf{\ }First, in Fig.
\ref{fig:Beta_only_graph} we display the stability map for the
exact soliton solutions (\ref{soliton}) -- (\ref{cos}) in the
plane $\left( \alpha _{1},q\right) $, with $\beta =0$ and $\alpha
_{0}=\left( \alpha _{0}\right) _{\mathrm{exact}}$, see Eq.
(\ref{alpha0}), and also for the solitons found numerically at the
same values of the parameters, except for the self-driving FF
coefficient, $\beta =0.1$. Note that the stability region for the
exact solitons is located inside the stripe corresponding to the
necessary stability condition (\ref{interval}), being actually
much narrower than it (which means that there are strong
nontrivial stability conditions for the soliton proper, which do
not amount to the simple criterion that guarantees the stability
of the zero background).

In Fig. \ref{fig:Beta_only_graph} (and similarly in Fig. \ref{fig:Alpha_90},
see below) we distinguish between stability sub-regions in which all the
eigenvalues are real, and those where complex ones are found. Accordingly, a
perturbation applied to the stable soliton in the latter sub-region excites
a damped intrinsic oscillatory mode (this will be observed below, in the
case of interactions between the solitons). Quite naturally, the complex
eigenvalues are found in the case when the loss parameter $\alpha _{1}$ is
small enough.

\begin{figure}[tbph]
\begin{center}
\includegraphics[width=2.5in]{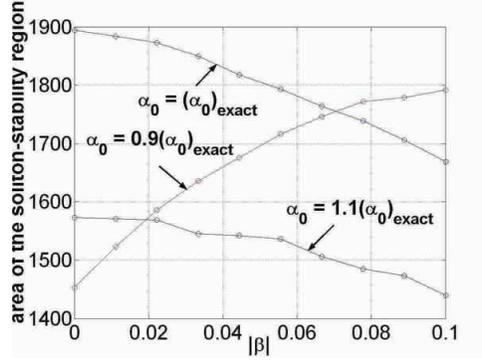}
\end{center}
\caption{The area of the stability region vs. $\protect\beta $, for three
different values of $\protect\alpha _{0}$. }
\label{Fig:area}
\end{figure}

A similar stability map is shown in Fig. \ref{fig:Alpha_90} for
the same case which was selected above for Fig. \ref{Fig:generic},
i.e., $\alpha _{0}=0.9\left( \alpha _{0}\right)
_{\mathrm{exact}}$. Additionally, for both cases $\alpha
_{0}=\left( \alpha _{0}\right) _{\mathrm{exact}}$ and $\alpha
_{0}=0.9\left( \alpha _{0}\right) _{\mathrm{exact}}$, as well as
for another one, $\alpha _{0}=1.1\left( \alpha _{0}\right)
_{\mathrm{exact}}$, Fig. \ref{Fig:area} shows the area of the
stability region as a function of $\beta $, between the two values
chosen for the display in Figs. \ref{fig:Beta_only_graph} and
\ref{fig:Alpha_90}, $\beta =0$ and $\beta =0.1$. Note that two
plots in Fig. \ref{Fig:area} pertain to exactly the same soliton
subfamilies which were included in Figs. \ref{Fig:generic0} and
\ref{Fig:generic}. A noticeable observation is that, depending on
the value of $\alpha _{0}$, the stability area may both decrease
and increase with $\beta $. As concerns the additional case of
$\alpha _{0}=1.1\left( \alpha _{0}\right) _{\mathrm{exact}}$,
included in Fig. \ref{Fig:area}, the stability map for it is not
very different from that for $\alpha _{0}=\left( \alpha
_{0}\right) _{\mathrm{exact}}$, see Fig.
\ref{fig:Beta_only_graph}, therefore it is not displayed here
separately.

\begin{figure}[tbph]
\begin{center}
\includegraphics[width=2.5in]{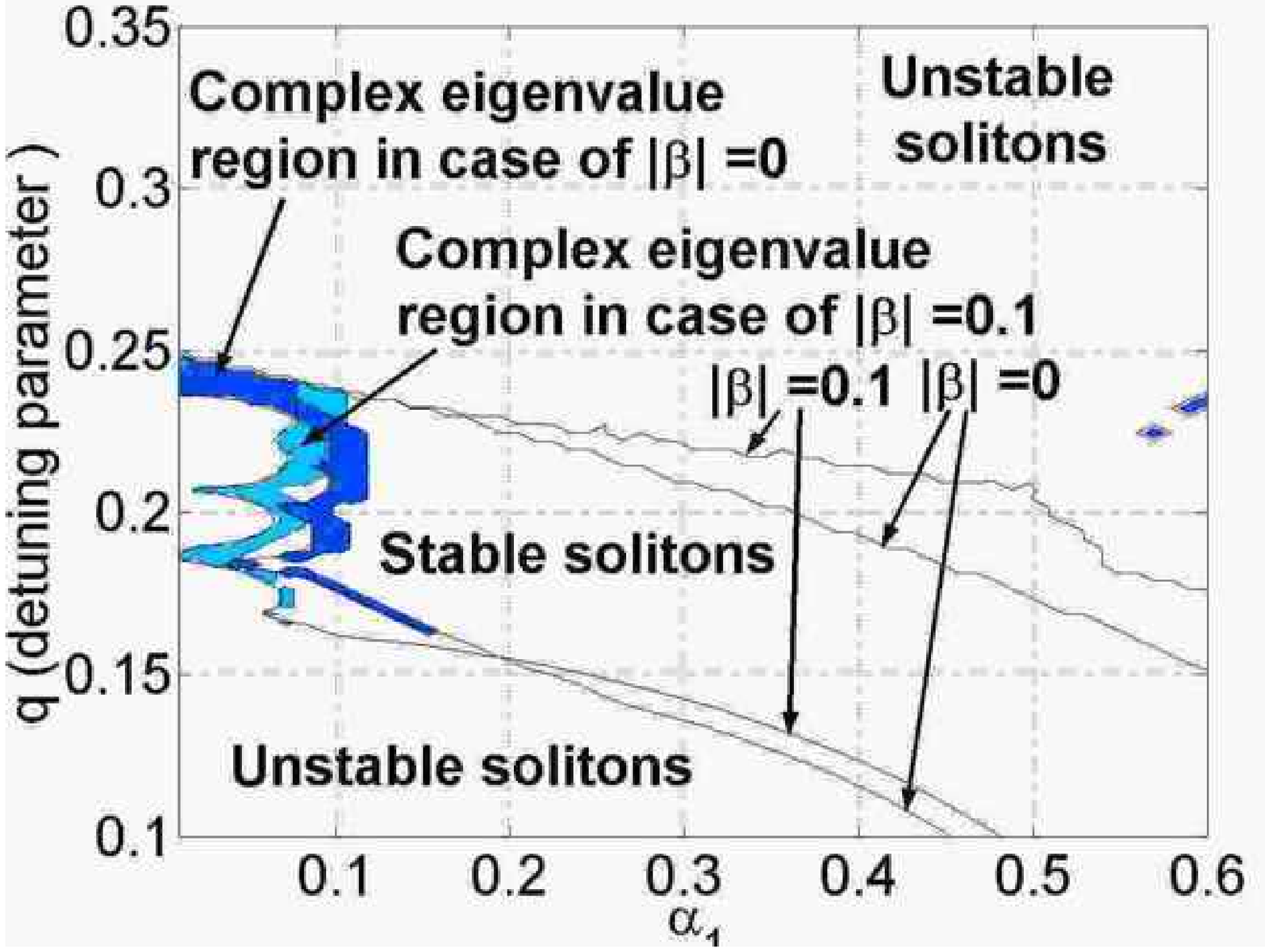}
\end{center}
\caption{The same as in Fig. \protect\ref{fig:Beta_only_graph},
but for $\protect\alpha _{0}=0.9\left( \protect\alpha _{0}\right)
_{\mathrm{exact}}$.} \label{fig:Alpha_90}
\end{figure}

\begin{figure}[tbph]
\centering
\includegraphics[width=3.5in]{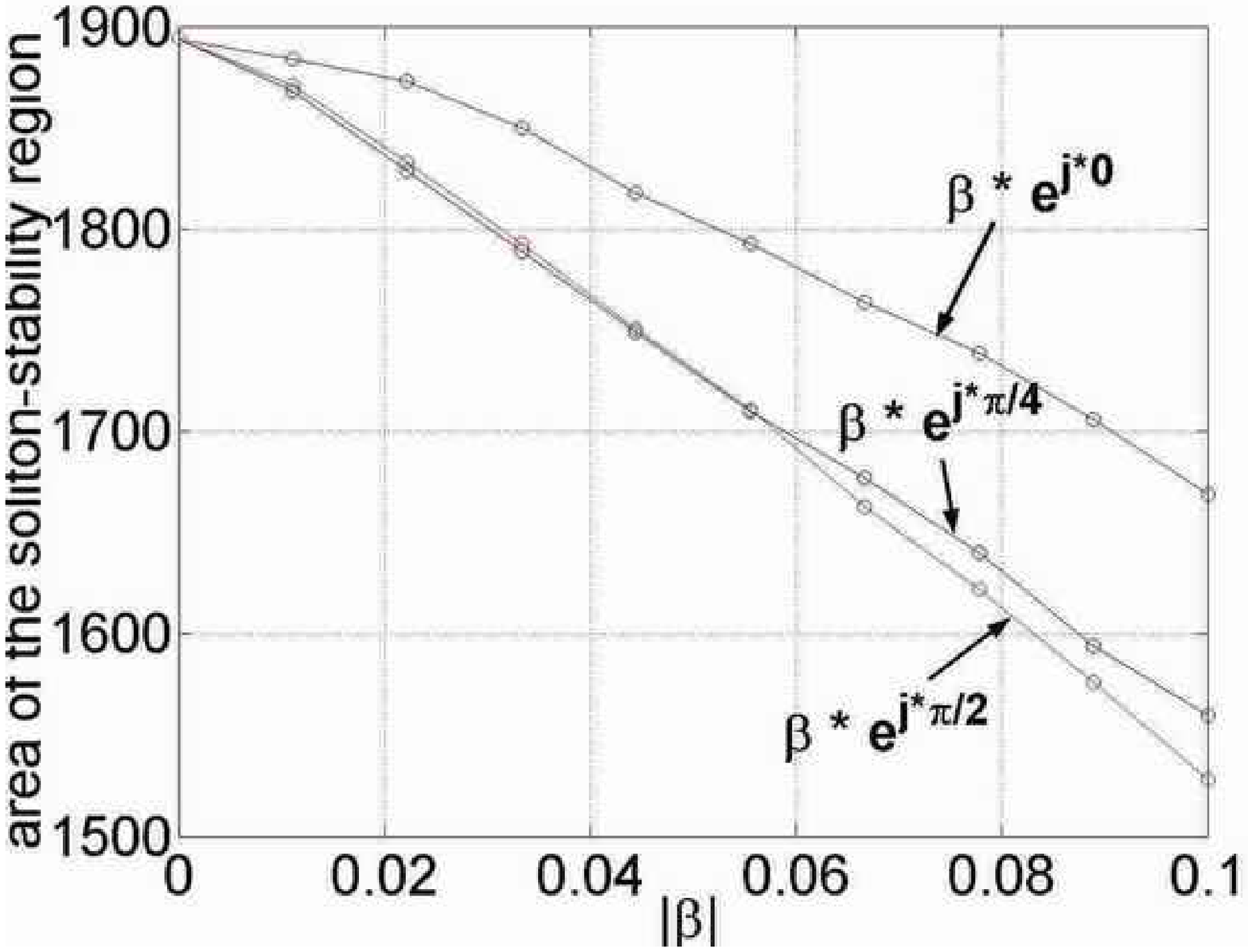}
\caption{The same as in Fig. \protect\ref{Fig:area} for
$\protect\alpha _{0}=\left( \protect\alpha _{0}\right)
_{\mathrm{exact}}$ and complex $\protect\beta $.}
\label{fig:Beta_phase_area_graph}
\end{figure}

We have also investigated the case when the self-driving
coefficient $\beta $ in the FF equation (\ref{u}) is complex. In
this case, the stability maps are not drastically different from
the ones displayed above for real $\beta $ (therefore we do not
show them here), although the area of the stability region gets
somewhat smaller. The area is shown, as a function of $|\beta |$,
for two different cases, $\beta =\left( 1+i\right) |\beta
|/\sqrt{2}$ and $\beta =i|\beta |$, in Fig.
\ref{fig:Beta_phase_area_graph}.

The conclusions concerning the stability of the solitons, which were drawn
above on the basis of the stability eigenvalues, where also verified against
direct simulations of the full system of Eqs. (\ref{u}) and (\ref{v}),
performed at a sufficiently dense grid of points covering the predicted
stability and instability regions in the system's phase space. As a result,
it has been concluded that all the solitons, which were predicted to be
stable, are stable indeed in the direct evolution. The solitons which are
expected to be unstable decay to zero under the action of small
perturbations or, if the perturbation is stronger, they may rearrange
themselves into solitons of a new kind, as described below.

\subsection{A bifurcation to the second type of stable solitons}

One may observe that the stability areas displayed in Figs.
\ref{fig:Beta_only_graph} and \ref{fig:Alpha_90} are located at
$q<0.25$. In particular, although we have no analytical proof of
the instability of the exact soliton (\ref{soliton}) at $q>0.25$,
we notice that the value of the mismatch parameter
$q=q_{\mathrm{c}}\equiv 0.25$ is a special one for the exact
solution, as Eqs. (\ref{cos}) show that the phase $\phi $ vanishes
precisely at $q=q_{\mathrm{c}}$.

\begin{figure}[tbph]
\begin{center}
$\begin{array}{c@{\hspace{0.5in}}c}
\includegraphics[width=2.5in]{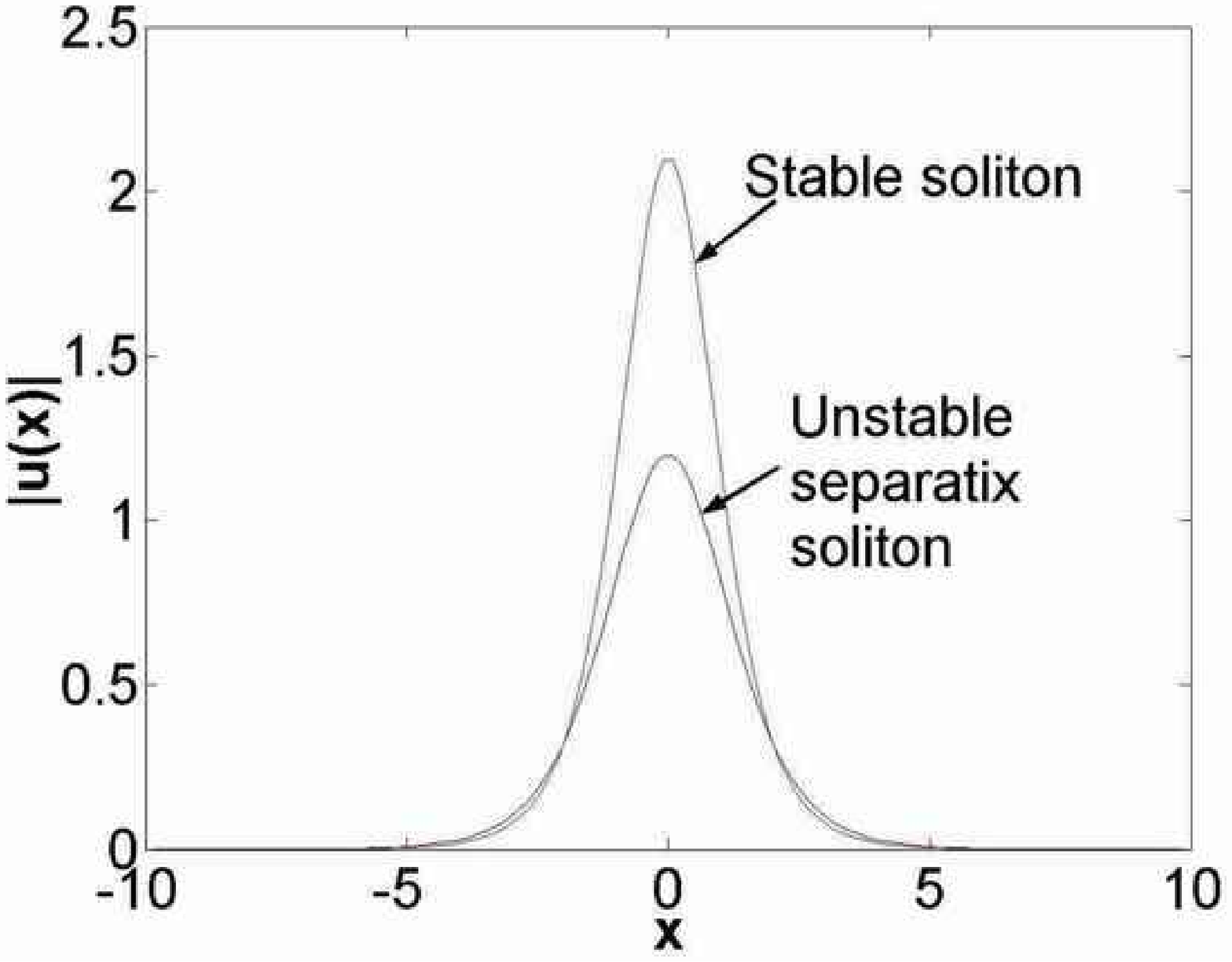} & \includegraphics[width=2.5in]{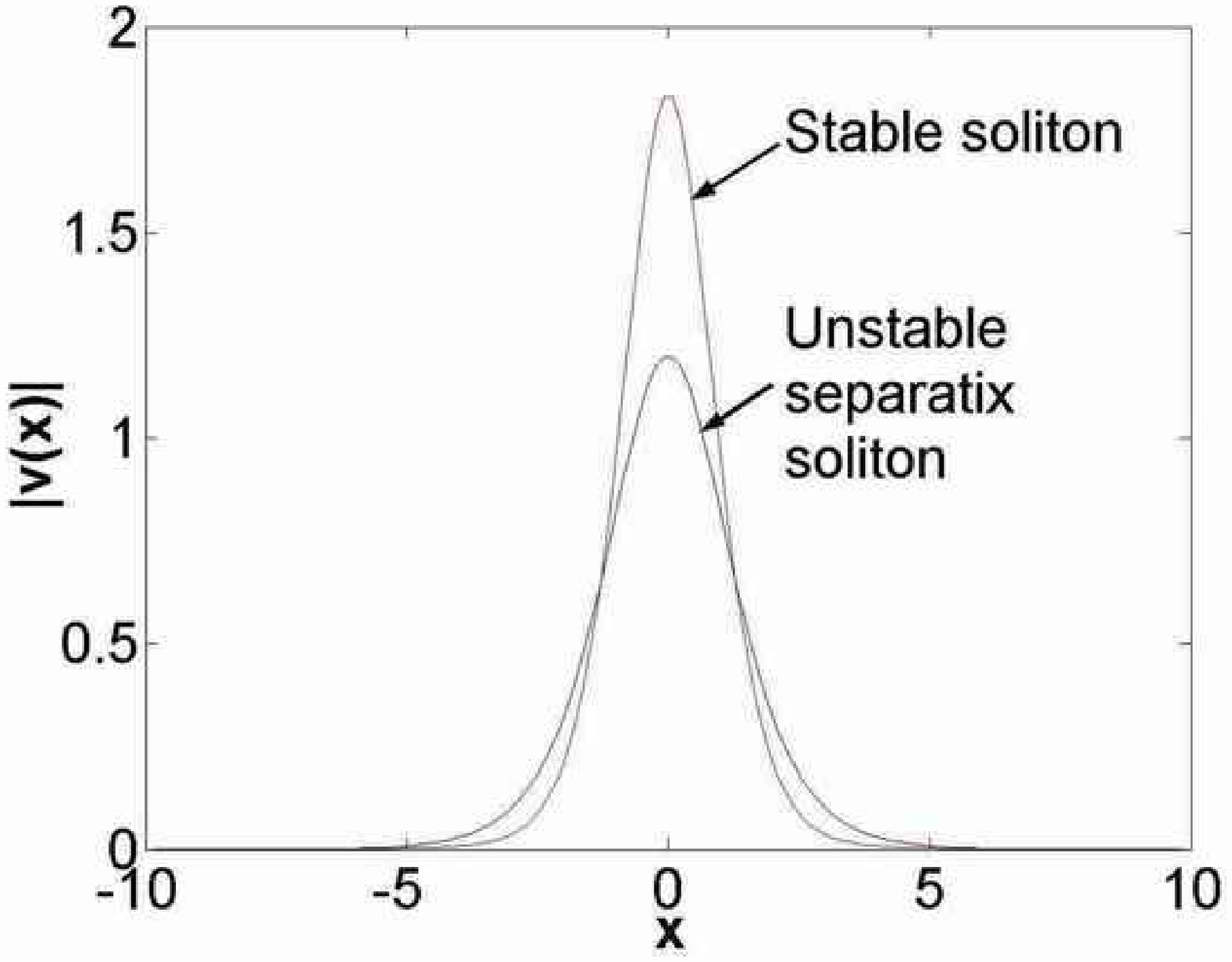} \\[0.4cm]
\mbox{\bf (a)} & \mbox{\bf (b)}\end{array}$\end{center}
\caption{The shape of the unstable type-I (separatrix) and stable
type-II solitons in the FH (a) and SH (b) components for for a
value of the mismatch parameter exceeding its critical value,
$q=0.3>q_{\mathrm{c}}\equiv 0.25$. The other parameters are
$\protect\beta =0$, $\protect\alpha _{1}=0.3$, and $\protect\alpha
_{0}=\left( \protect\alpha _{0}\right) _{\mathrm{exact}}\approx
0.36$, as per Eq. (\protect\ref{alpha0}).}
\label{Fig:separatrix_shape}
\end{figure}

A stable soliton of a different type can be found at $q>0.25$.
This solution, which we will call a type-II soliton, to
distinguish it from the one considered above, that we will refer
to as type-I, cannot be obtained by the continuation of the exact
solution (\ref{soliton}). This makes finding the type-II soliton
directly from the numerical solution of the stationary version of
Eqs. (\ref{FF}) and (\ref{SH}) problematic, as a good initial
guess is not available. Nevertheless, it was found, in the region
$q>0.25$, in the following way: one can take the numerically exact
unstable type-I soliton, and add an arbitrary perturbation to it.
A small perturbation initiates decay of the unstable soliton to
zero. However, if the perturbation is sufficiently large, the
outcome may be spontaneous rearrangement of the pulse into a new
stable one, which we identify as the type-II soliton. An example
of the unstable type-I soliton (alias \textit{separatrix soliton},
see below), and its stable type-II counterpart, which is generated
from it by a perturbation, is displayed in Fig.
\ref{Fig:separatrix_shape}.

\begin{figure}[tbph]
\centering
\includegraphics[width=3.5in]{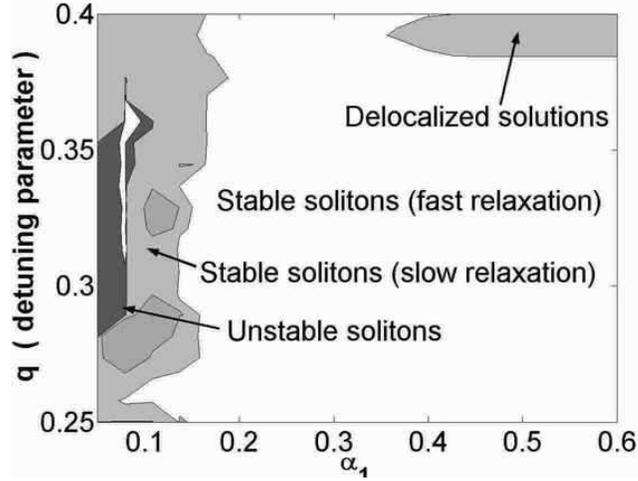}
\caption{The stability map for the type-II soliton self-trapping
from the unstable separatrix (type-I) soliton in the region
$q>0.25$. In the regions marked as ``fast" and ``slow" relaxation,
the unstable soliton directly relaxes into the stable one, or does
so via formation of a breather with slowly decaying intrinsic
oscillations. In the region of ``unstable solitons", the unstable
soliton always decays to zero, and in the region of ``delocalized
solutions", a spatially periodic state sets in, instead of a
stable soliton. The map pertains to $\protect\beta =0$, with
$\protect\alpha _{0}$ chosen as per Eq. (\protect\ref{alpha0}). }
\label{fig:separatrix_to_stable_solitons}
\end{figure}

The stability map for the type-II solitons was identified, as it
was done above for their type-I counterparts in the region
$q<0.25$, both by means of the computation of the eigenvalues from
the linearized equations, and by means of direct simulations. The
resulting map is displayed in Fig.
\ref{fig:separatrix_to_stable_solitons}. It shows not only the
region where the type-II soliton is stable, but also indicates if
the perturbation-induced evolution of the unstable type-I solution
leads directly to the establishment of the stable type-II soliton,
or, instead, transition to an excited state (``breather") is
observed, which then slowly relaxes into the stable type-II
soliton. Also shown are regions where the stable soliton does not
self-trap; instead, the unstable soliton either decays to zero, or
generates a stable delocalized state.

The solitons of types I and II are actually connected by a bifurcation.
Indeed, the model which supports a stable soliton is a bistable system, as
the zero solution is stable too in this case. According to well-known
general principles, in a bistable system a separatrix must exist, which is a
border between attraction basins of the two stable states. Usually, the role
of the separatrix is played by an extra unstable soliton solution, whose
amplitude is smaller and width larger than those of the stable soliton (see,
e.g., Refs. \cite{PhysicaD} and \cite{Barash}). The unstable type-I soliton
is such a separatrix in the case of $q>q_{\mathrm{c}}\equiv 0.25$.

\begin{figure}[tbph]
\centering\includegraphics[width=3.5in]{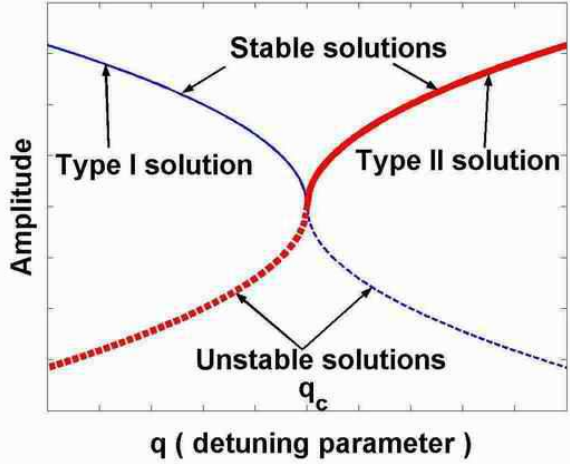}
\caption{The assumed stability-swap bifurcation between the solitons of
types I and II (thin and bold curves) which occurs at the critical value of
the mismatch, $q=q_{\mathrm{c}}=0.25$. Stable and unstable solutions are
depicted by solid and dashed lines, and the vertical axis refers to the
solitons' amplitudes.}
\label{fig:bifurcation}
\end{figure}

The fact that the type-I soliton is stable at $q<q_{\mathrm{c}}$,
while no stable soliton of type II was found in that region,
implies that a stability-swap bifurcation, involving both
solitons, occurs at $q=$ $q_{\mathrm{c}}$. A conjectured
bifurcation diagram is schematically depicted in Fig.
\ref{fig:bifurcation}. \textbf{\ }

Note that the bifurcation diagram includes an unstable branch in the region
of $q<q_{\mathrm{c}}$, which is a conjectured continuation of the type-II
solution to $q<q_{\mathrm{c}}$. This branch cannot be readily found from the
stationary version of Eqs. (\ref{FF}) and (\ref{SH}), since a good initial
guess is not available for it, and it cannot be found in dynamical
simulations either, as it is unstable. Thus, while we did not make a strong
effort to explicitly find this branch -- as, being unstable, it has no
direct physical significance -- we assume that such a branch exists.

\subsection{Stability of solitons in a generalized model}

The model based on Eqs. (\ref{u}) and (\ref{v}) can be generalized
by replacing the evolution and spatial variables $t$ and $x$,
respectively, by $z$ and $\tau $:\begin{eqnarray}
iu_{z}+\frac{1}{2}u_{\tau \tau }+u^{\ast }v &=&\left( 1-i\alpha
_{1}\right)
u+i\alpha _{0}v^{\ast }+i\beta \left( u^{\ast }\right) ^{2},  \nonumber \\
iv_{z}+\frac{D}{4}v_{\tau \tau }+\frac{1}{2}u^{2} &=&\left( 2q-i\alpha
_{1}\right) v+i\alpha _{0}u^{\ast }.  \label{tau}
\end{eqnarray}This model corresponds, instead of the spatial solitons in cavities, to
temporal solitons in waveguides. The temporal solitons are localized in the
reduced-time variable $\tau $, and propagate along the coordinate $z$. In
Eqs. (\ref{tau}), the second derivatives account for the temporal dispersion
[rather than diffraction, in the original model (\ref{u}), (\ref{v})], $D$
being a relative SH/FF dispersion coefficient. We have checked that, in this
generalization (for instance, with $D=2$, instead of $D=1$), the results for
the shape and stability of solitons are very similar to those reported above.

\subsection{\textbf{Investigation of moving solitons}}

A straightforward extension of the above analysis is to search for
moving solitons. Their existence in the present model is not
obvious, as the drive terms in Eqs. (\ref{u}) and (\ref{v})
(unlike the loss terms) break the Galilean invariance of the
equations. Without the drive, the Galilean boost with an arbitrary
speed $C$,\begin{equation} u\rightarrow u\cdot
e^{iCX},v\rightarrow v\cdot e^{2iCX},  \label{boost}
\end{equation}transforms any solution $u(x),v(x)$ into a moving one, $u(x-Ct),v(x-Ct)$.

\begin{figure}[tbph]
\begin{center}
$\begin{array}{c@{\hspace{0.5in}}c}
\includegraphics[width=2.5in]{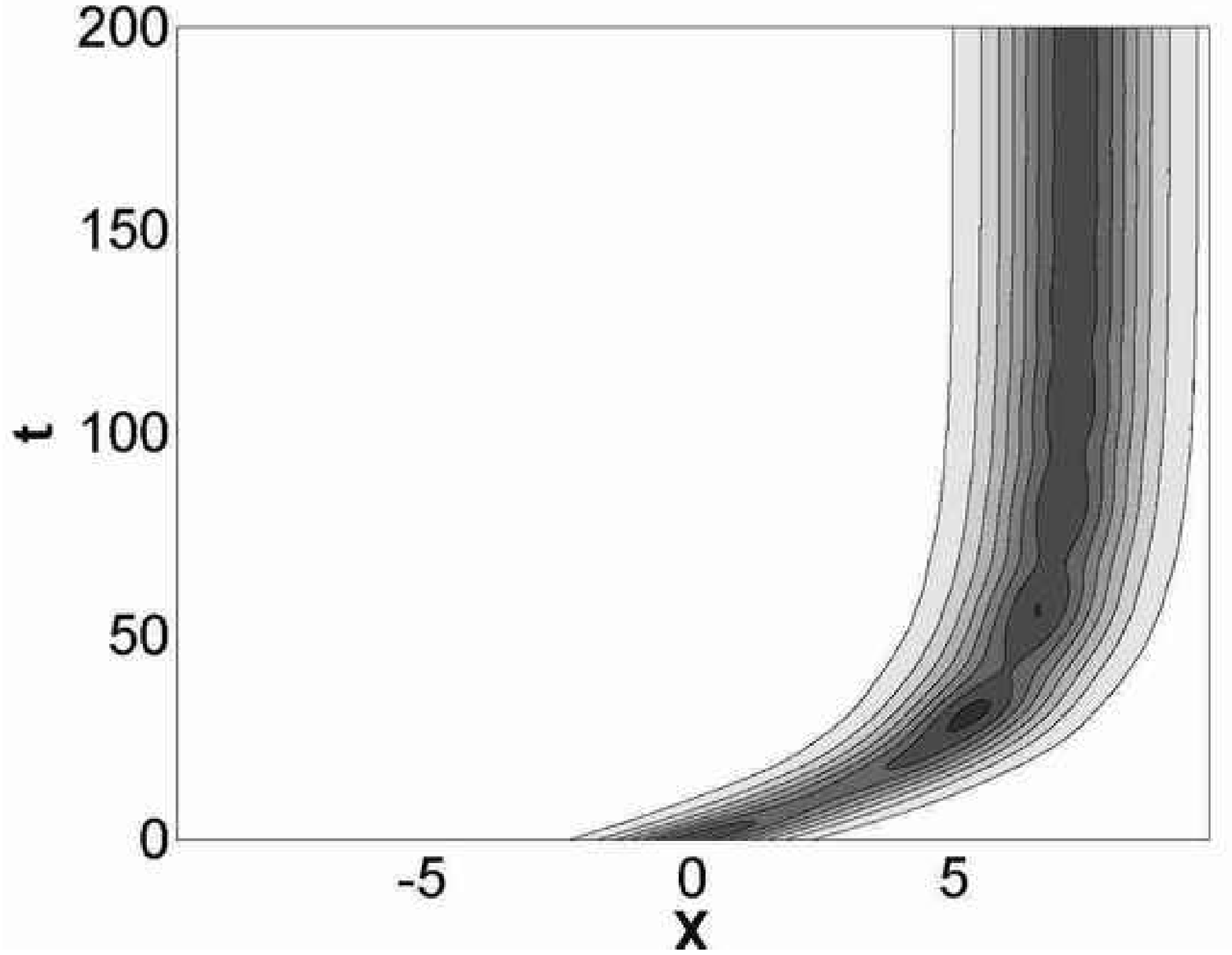} & \includegraphics[width=2.5in]{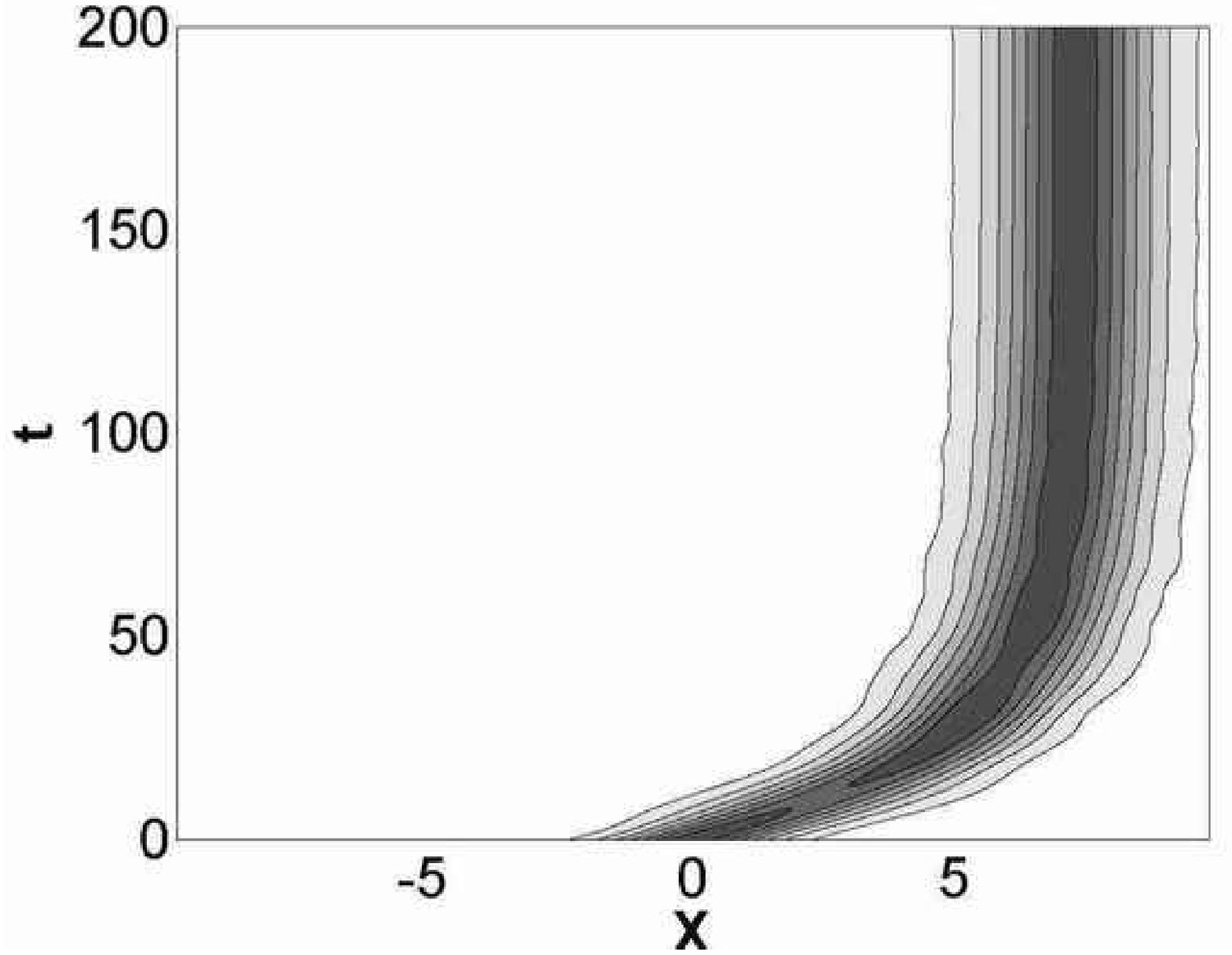} \\[0.4cm]
\mbox{\bf (a)} & \mbox{\bf (b)}\end{array}$\end{center}
\caption{An example of motion and subsequent stoppage of the
boosted soliton (\protect\ref{soliton}), in the case of
$\protect\alpha _{1}=0.02$, $q=0.245$, $\protect\beta =0$,
$\protect\alpha _{0}=\left( \protect\alpha _{0}\right)
_{\mathrm{exact}}=2.828$ [see Eq. (\protect\ref{alpha0})], and
$C=0.3$ (this value of the boost is close to the critical value,
which is $C_{\mathrm{cr}}=0.3335$ in the present case). The panels
(a) and (b) show the evolution of the pulse's amplitude profiles
in the FF and SH components, respectively.}
\label{Fig:exact_Galilean_stop}
\end{figure}

\begin{figure}[tbph]
\begin{center}
$\begin{array}{c@{\hspace{0.5in}}c}
\includegraphics[width=2.5in]{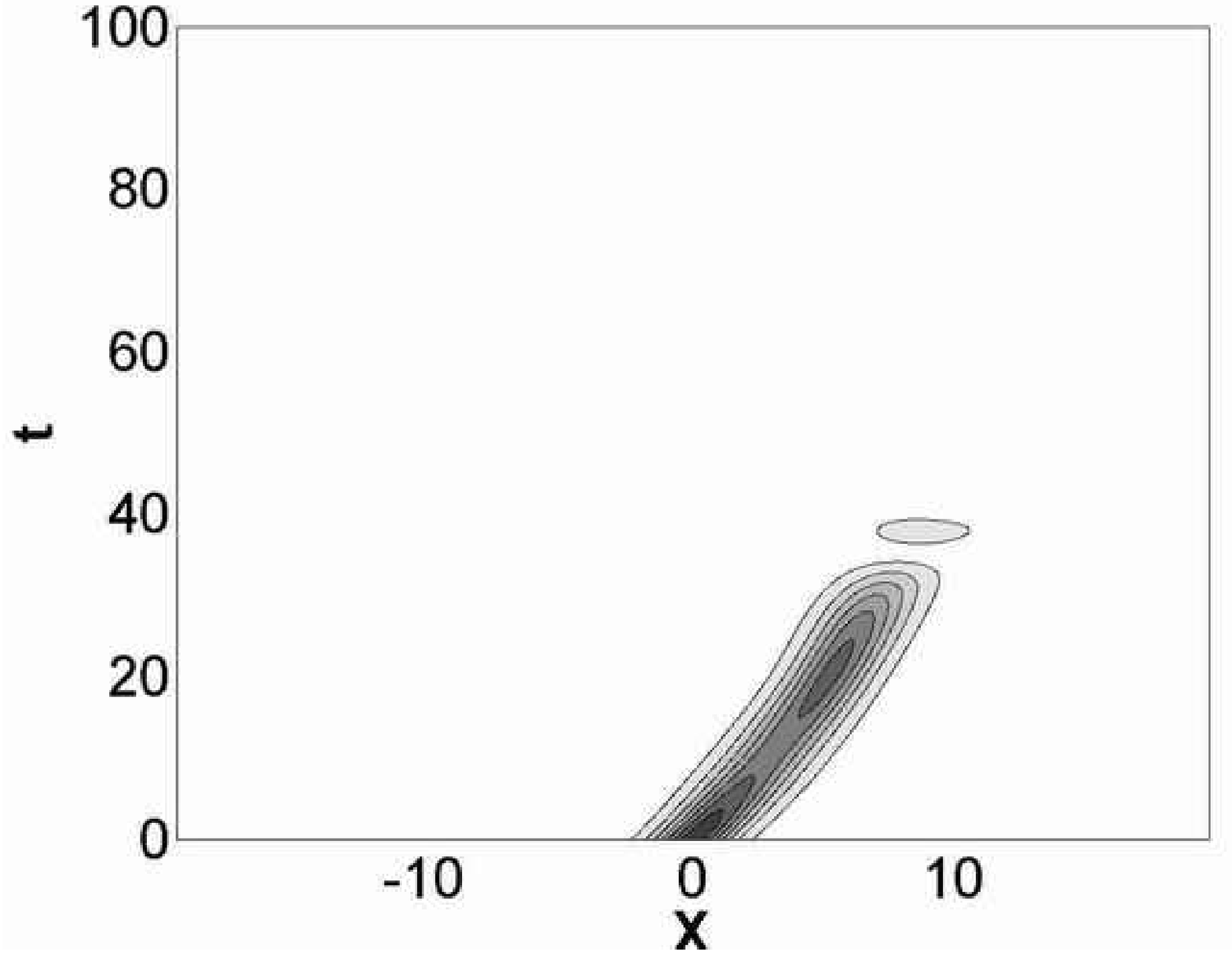} & \includegraphics[width=2.5in]{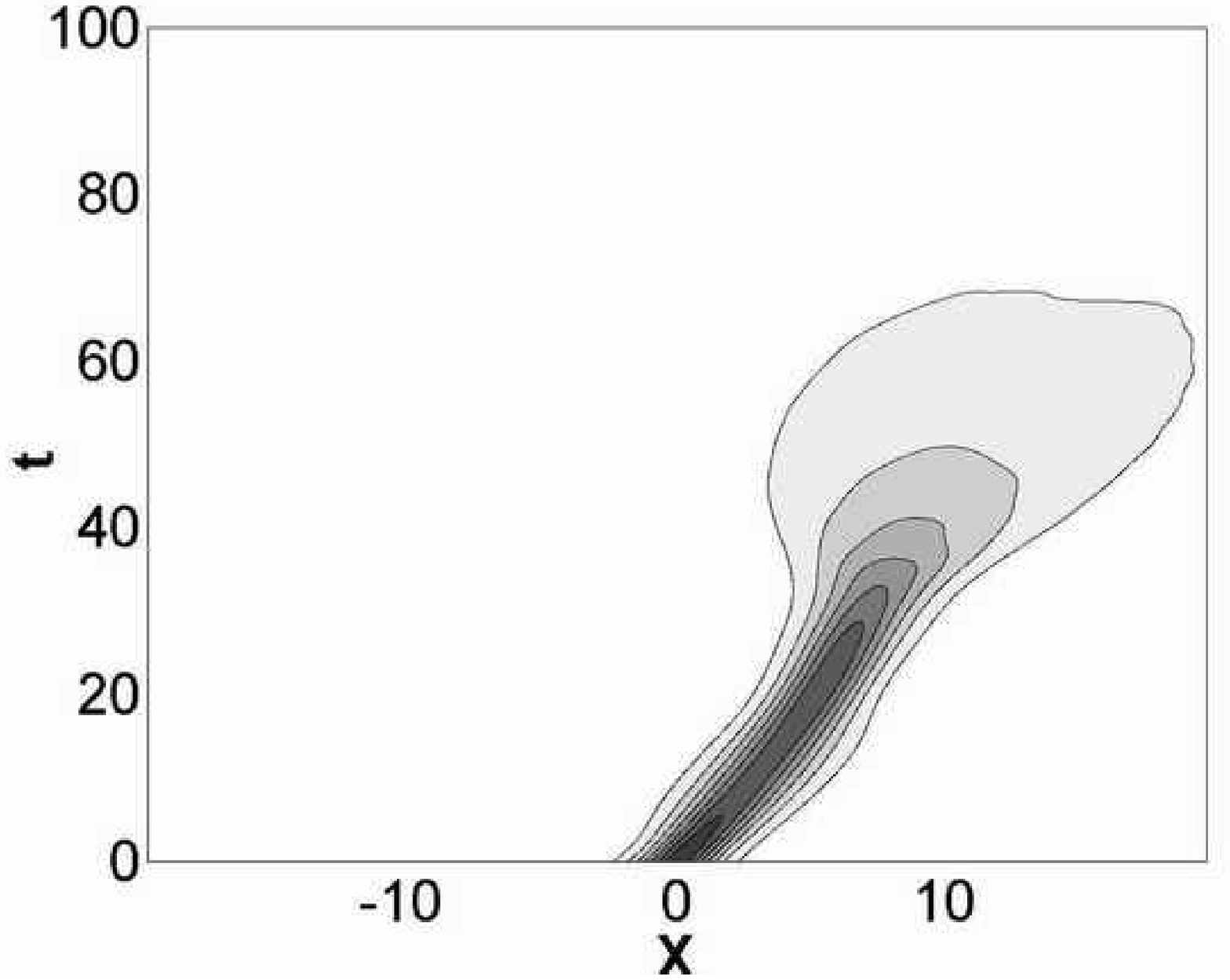} \\[0.4cm]
\mbox{\bf (a)} & \mbox{\bf (b)}\end{array}$\end{center}
\caption{The same as in Fig.
\protect\ref{Fig:exact_Galilean_destroy}, but with the boost
parameter $C=0.35$ slightly exceeding the critical value
$C_{\mathrm{cr}}$.} \label{Fig:exact_Galilean_destroy}
\end{figure}

To investigate the possibility of the existence of moving solitons
in the present model, we ran systematic numerical experiments,
taking the stable soliton solutions in the analytical or numerical
form, as found above, and simulating the full system of Eqs.
(\ref{u}) and (\ref{v}) with the boosted initial conditions
(\ref{boost}). The result is that steadily moving solitons are not
possible. Instead, a critical value $C_{\mathrm{cr}}$ of the boost
parameter $C$ was found, such that the soliton moves for a while
but quickly comes to a halt and returns to its initial form, if
$C<C_{\mathrm{cr}}$, as shown in Fig.
\ref{Fig:exact_Galilean_stop}. In the opposite case,
$C>C_{\mathrm{cr}}$, the soliton always gets destroyed, see an
example in Fig. (\ref{Fig:exact_Galilean_destroy}).

\begin{figure}[tbph]
\begin{center}
\includegraphics[width=2.5in]{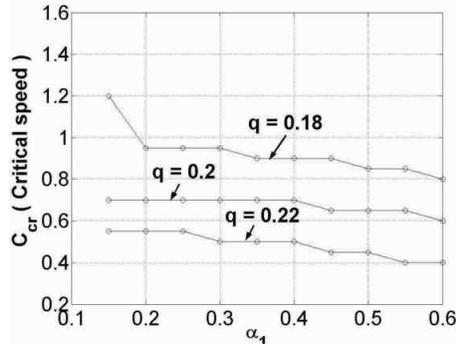}
\end{center}
\caption{The critical value $C_{\mathrm{cr}}$ of the boost
parameter (initial speed), which is defined in Eqs.
(\protect\ref{boost}), as a function of $\protect\alpha _{1}$ at
fixed values of $q$, for the exact solitons
(\protect\ref{soliton}) [i.e., with $\protect\beta =0$, and
$\protect\alpha _{0}=\left( \protect\alpha _{0}\right)
_{\mathrm{exact}}$]. The soliton comes to a halt, or gets
destroyed, respectively, in the cases $C<C_{\mathrm{cr}}$ and
$C>C_{\mathrm{cr}}$, as shown in Figs.
\protect\ref{Fig:exact_Galilean_stop} and
\protect\ref{Fig:exact_Galilean_destroy}. }
\label{Fig:Critical_Speed}
\end{figure}

The critical boost $C_{\mathrm{cr}}$ depends on values of the
model's parameters. In particular, for the family of the exact
solitons taken in the form of Eqs. (\ref{soliton}), with $\alpha
_{0}$ selected as per Eq. (\ref{alpha0}), $C_{\mathrm{cr}}$ is
shown, as a function of the loss parameter $\alpha _{1}$ at
different fixed values of the mismatch $q$, in Fig.
(\ref{Fig:Critical_Speed}).

\begin{figure}[tbph]
\begin{center}
$\begin{array}{c@{\hspace{0.5in}}c}
\includegraphics[width=2.5in]{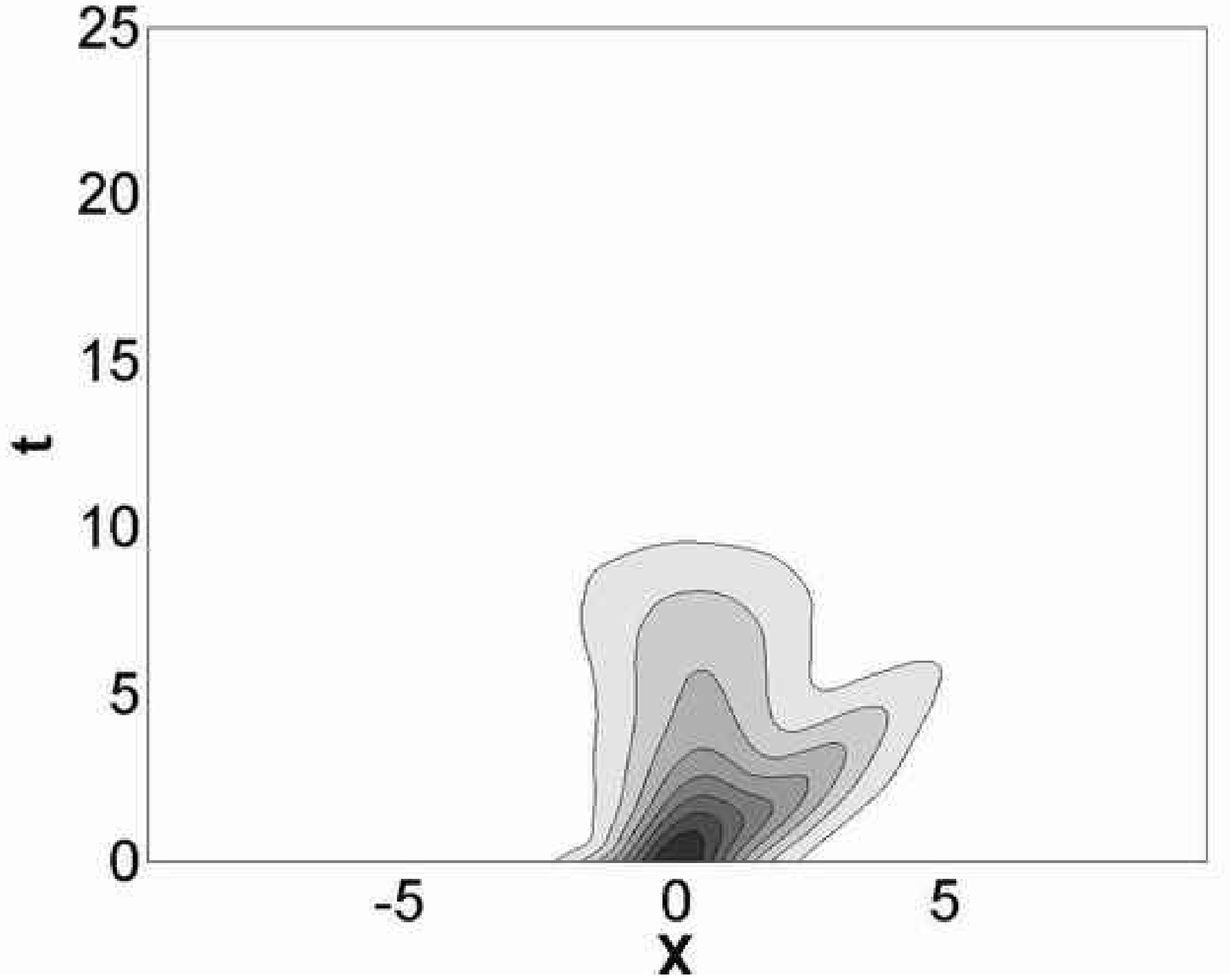} & \includegraphics[width=2.5in]{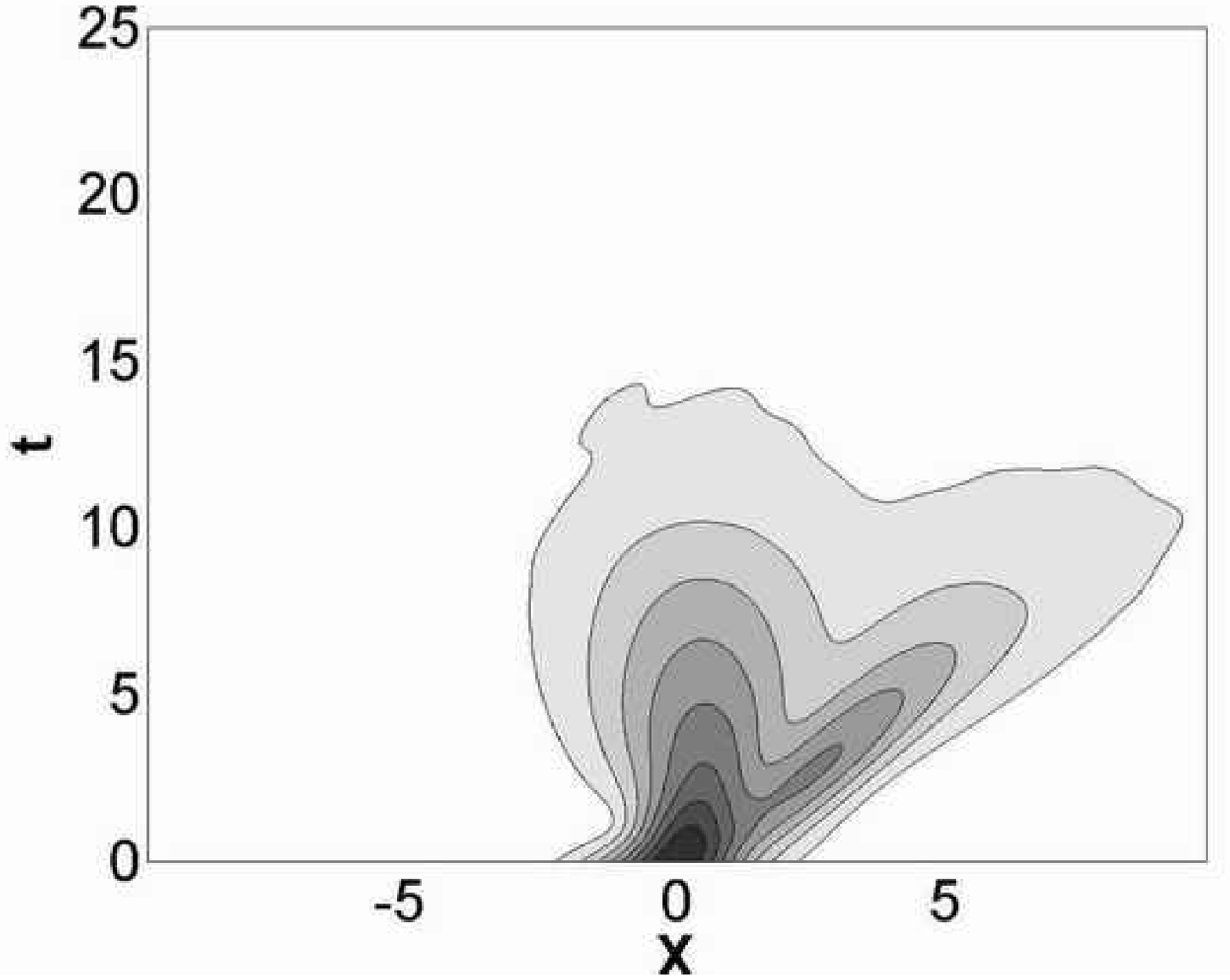} \\[0.4cm]
\mbox{\bf (a)} & \mbox{\bf (b)}\end{array}$\end{center}
\caption{An example of the splitting of the soliton
(\protect\ref{soliton}) to which an overcritical boost was
applied, and subsequent decay of the secondary pulses. In this
case, $\protect\alpha _{1}=0.15$, $q=0.2$, $\protect\beta =0$,
$\protect\alpha _{0}=\left( \protect\alpha _{0}\right)
_{\mathrm{exat}}=\allowbreak 0.25$, the critical value of the
boost is $C_{\mathrm{cr}}\approx 0.7$, and the actual value of the
boost is $C=0.75$. The panels (a) and (b) show the evolution of
the amplitude profiles in the FF and SH components, respectively.}
\label{Fig:splitting}
\end{figure}

The particular mode of the destruction of the boosted soliton in
the case of $C>C_{\mathrm{cr}}$ depends on values of the
parameters. In particular, instead of the straightforward decay,
as in Fig. \ref{Fig:exact_Galilean_destroy}, the soliton may first
split into two secondary pulses with different velocities, both of
which eventually decay. An example of the latter scenario is
displayed in Fig. \ref{Fig:splitting}. Also possible are
situations in which the soliton does not split, but its FF
component decays much faster than the SH one (although the FF and
SH loss parameters are equal).

\section{\textbf{Interactions between solitons}}

Once stable solitons have been found, the next necessary step in
the investigation of their fundamental dynamical properties, as
well as in the development of potential applications, is the study
of interactions between them. To this end, we ran systematic
simulations of configurations initially composed of two identical
stable solitons with centers placed at a distance $X_{0}$. Note
that, unlike the situation in models of the Ginzburg-Landau type
(the ones with the intrinsic gain), in models with the parametric
gain the phase of each individual soliton is locked to a single
value [see Eqs. (\ref{soliton}) and (\ref{cos})], hence the
relative phase of the two solitons is not a free parameter, but is
equal to zero \cite{Cai}; for this reason, the solitons always
attract each other. The simulations demonstrate that, in all the
cases, the attraction gives rise to merger of the two solitons
into a single one. If the loss parameter $\alpha _{1}$ is large
enough, so that the soliton existing at this value of $\alpha
_{1}$ has no complex eigenvalue of small intrinsic perturbations
(see Figs. \ref{fig:Beta_only_graph} and \ref{fig:Alpha_90}), the
resulting single soliton emerges in its stationary form. On the
other hand, if $\alpha _{1}$ is small, and the soliton possesses
an intrinsic complex eigenvalue, the final soliton appears in an
excited (vibrating) state, which then slowly relaxes to the static
one. The time necessary for the fusion of the two solitons into
one depends on the initial separation $X_{0}$, but the outcome of
the interaction does not depend on $X_{0}$.

Below, we illustrate these conclusions by typical examples. In all the
cases, we used exact soliton solutions (\ref{soliton}) to construct the
initial configuration. Simulations of the configuration composed of two
solitons of a more general form produced virtually the same results as those
displayed below.

\begin{figure}[tbph]
\begin{center}
$\begin{array}{c@{\hspace{0.5in}}c}
\includegraphics[width=2.5in]{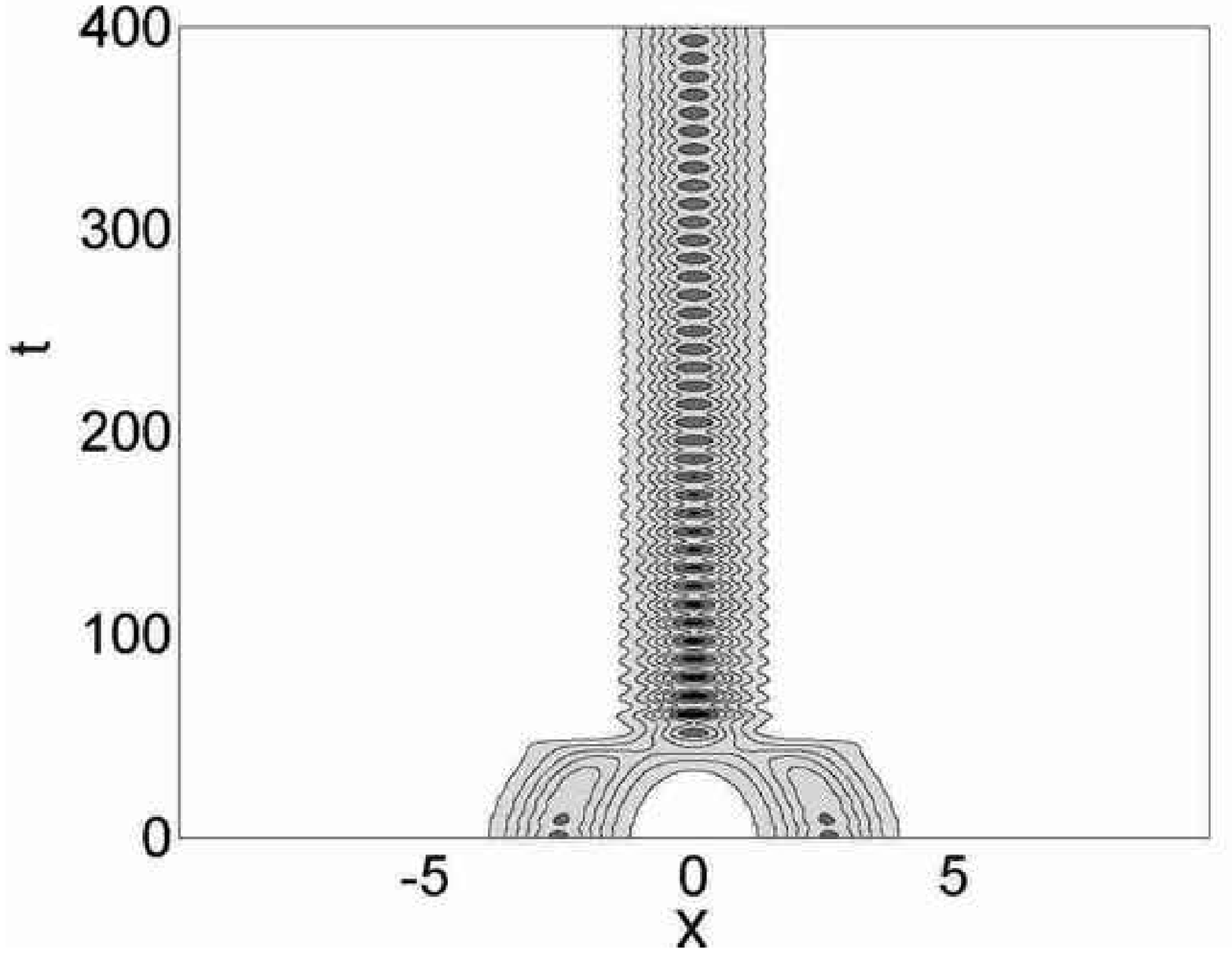} & \includegraphics[width=2.5in]{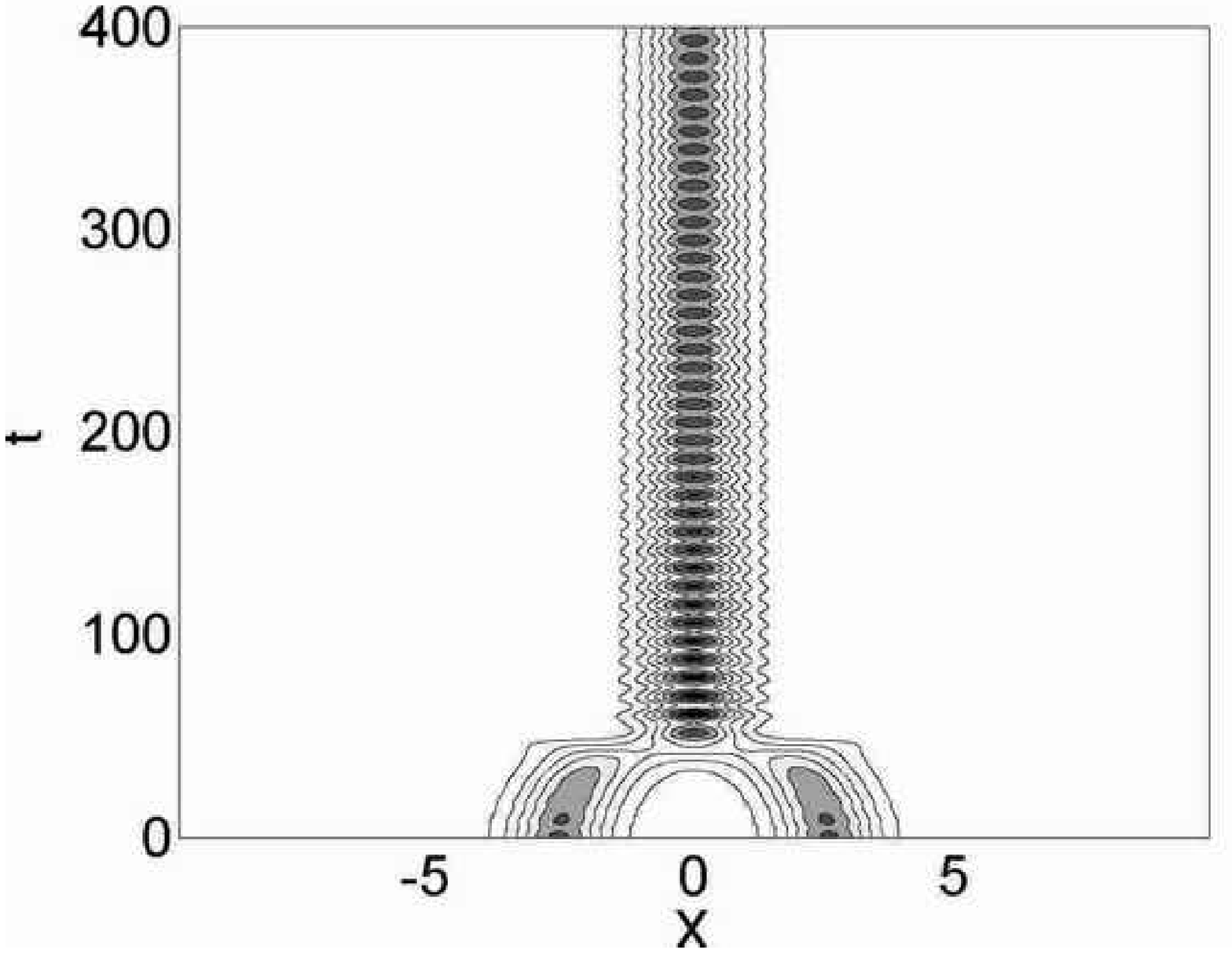} \\[0.4cm]
\mbox{\bf (a)} & \mbox{\bf (b)}\end{array}$\end{center}
\caption{The merger of two solitons, with the initial separation
$X_{0}=5.2$ between them, into a pulse with excited internal
vibrations, which later slowly relaxes into the stationary
soliton. The panels (a) and (b) show the FF and SH components of
the field, respectively. The parameters are $q=0.207$,
$\protect\alpha _{1}=0.094$, and $\protect\alpha _{0}=\left(
\protect\alpha _{0}\right) _{\mathrm{exact}}=0.196$, see Eq.
(\protect\ref{alpha0}).} \label{fig:Double_sol_3}
\end{figure}

\begin{figure}[tbph]
\begin{center}
$\begin{array}{c@{\hspace{0.5in}}c}
\includegraphics[width=2.5in]{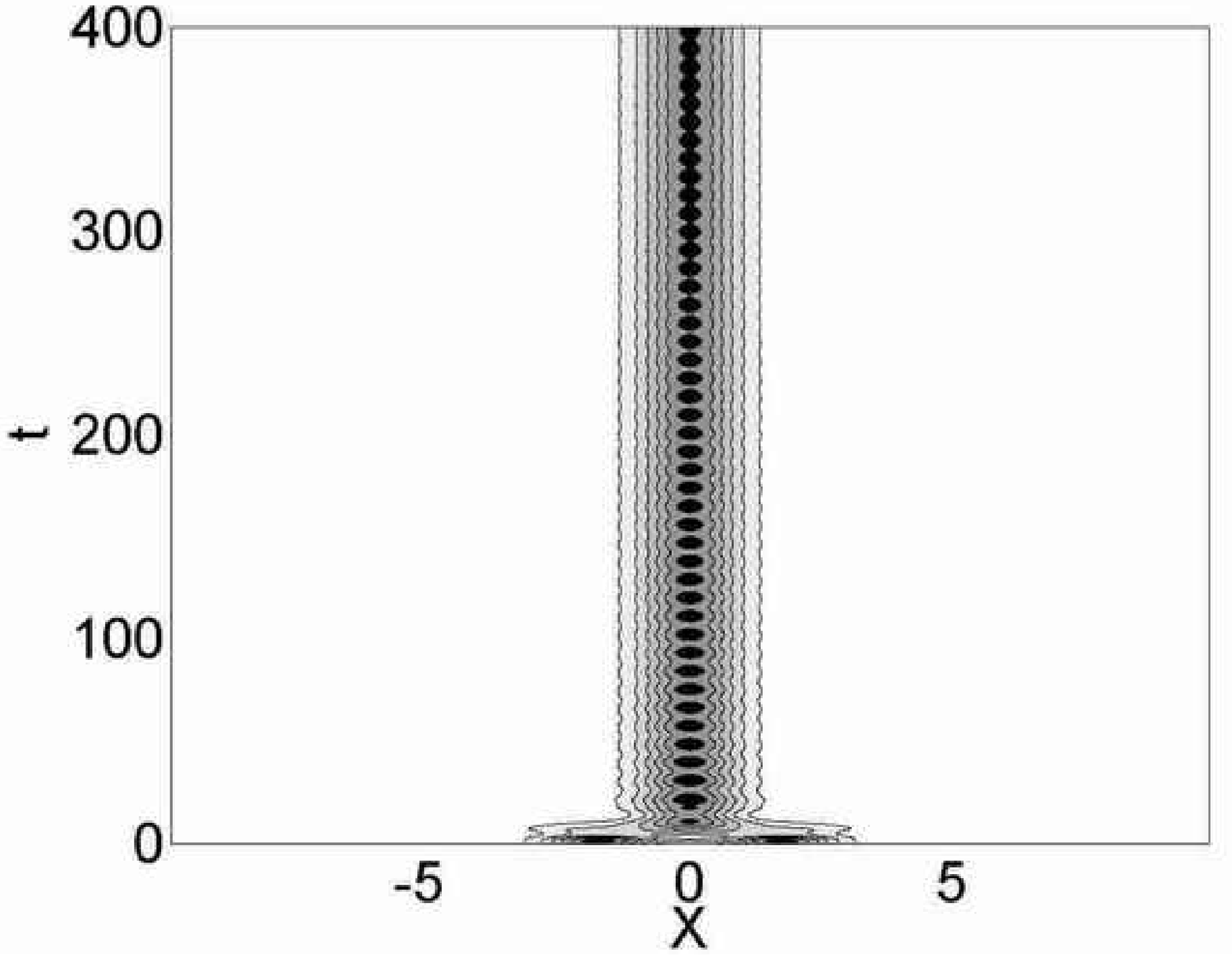} & \includegraphics[width=2.5in]{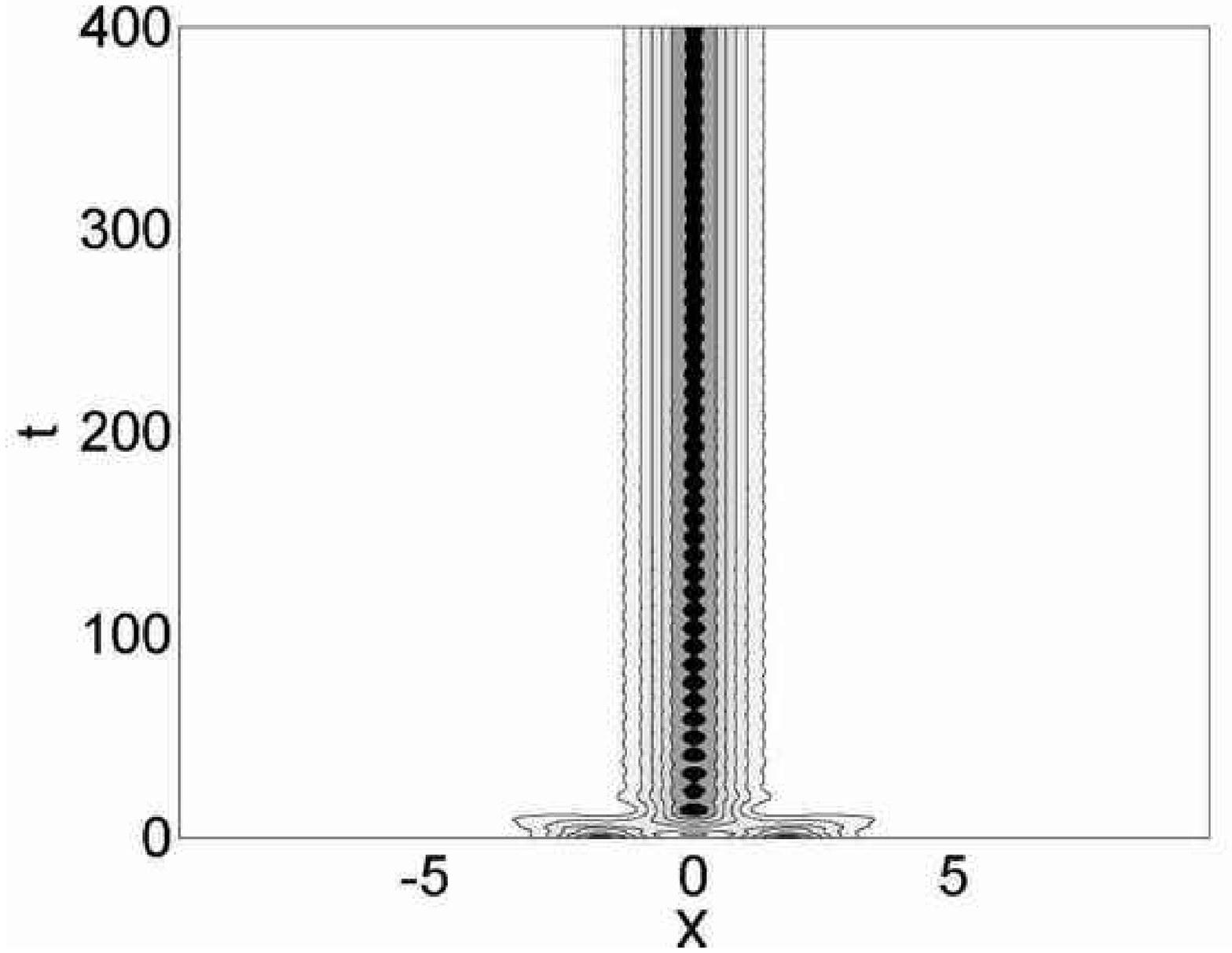} \\[0.4cm]
\mbox{\bf (a)} & \mbox{\bf (b)}\end{array}$\end{center}
\caption{The same as in Fig. \protect\ref{fig:Double_sol_3}, but
for a smaller initial separation between the solitons,
$X_{0}=3.6$. In this case, it is obvious that the vibrating pulse,
produced by the merger of the two initial solitons, relaxes
towards the static soliton.} \label{fig:Double_sol_4}
\end{figure}

If the loss parameter $\alpha _{1}$ is small enough, a pair of
exact solitons merge into a pulse which demonstrates nearly
persistent intrinsic vibrations, as it was said above and is shown
in Fig. \ref{fig:Double_sol_3}. Continuing the simulations on a
much longer time scale demonstrates that the vibrations slowly
fade out, and the pulse relaxes to the static configuration. The
latter feature is seen clearer in Fig. \ref{fig:Double_sol_4},
which corresponds to a smaller initial separation between the
solitons.

\begin{figure}[tbph]
\begin{center}
$\begin{array}{c@{\hspace{0.5in}}c}
\includegraphics[width=2.5in]{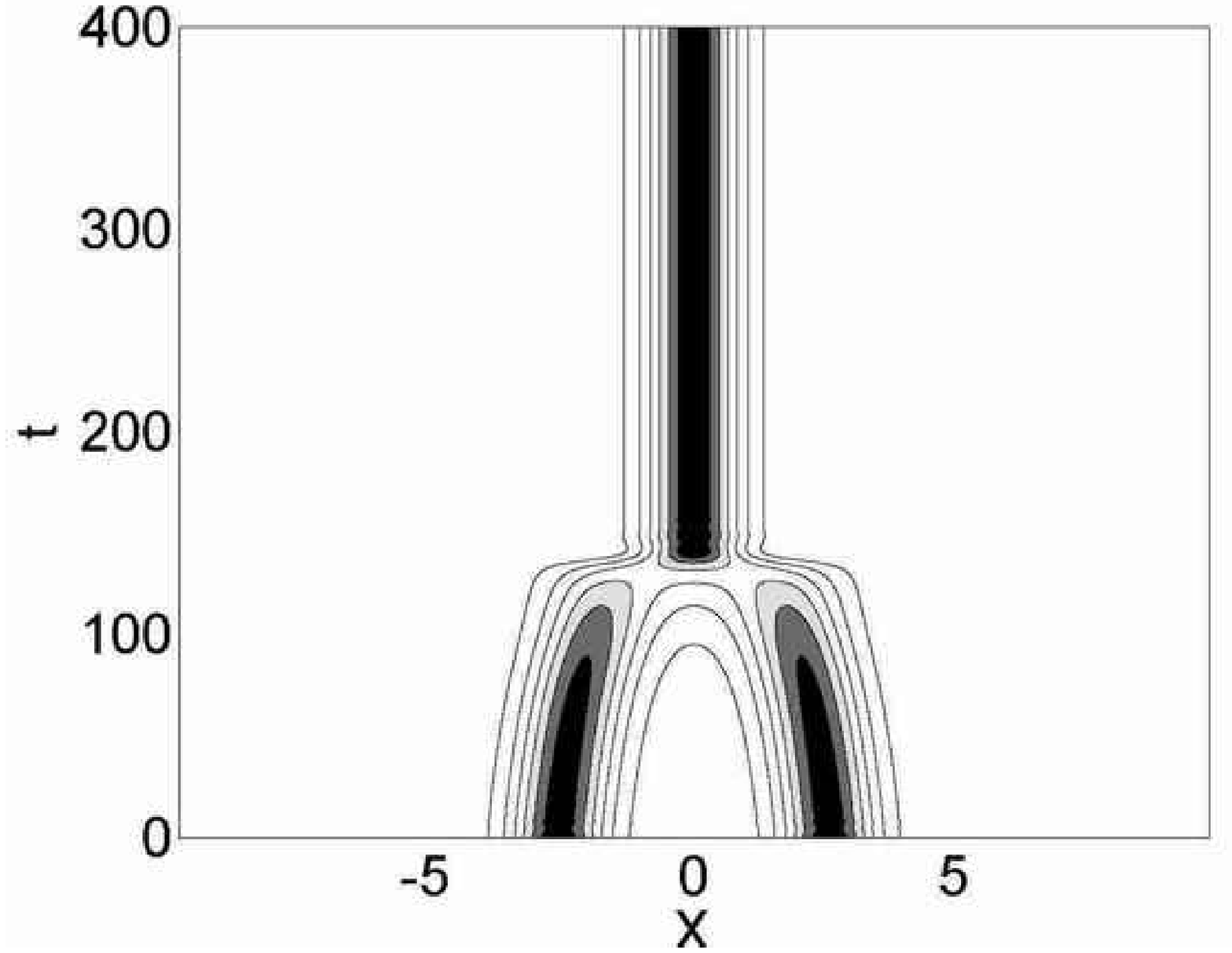} & \includegraphics[width=2.5in]{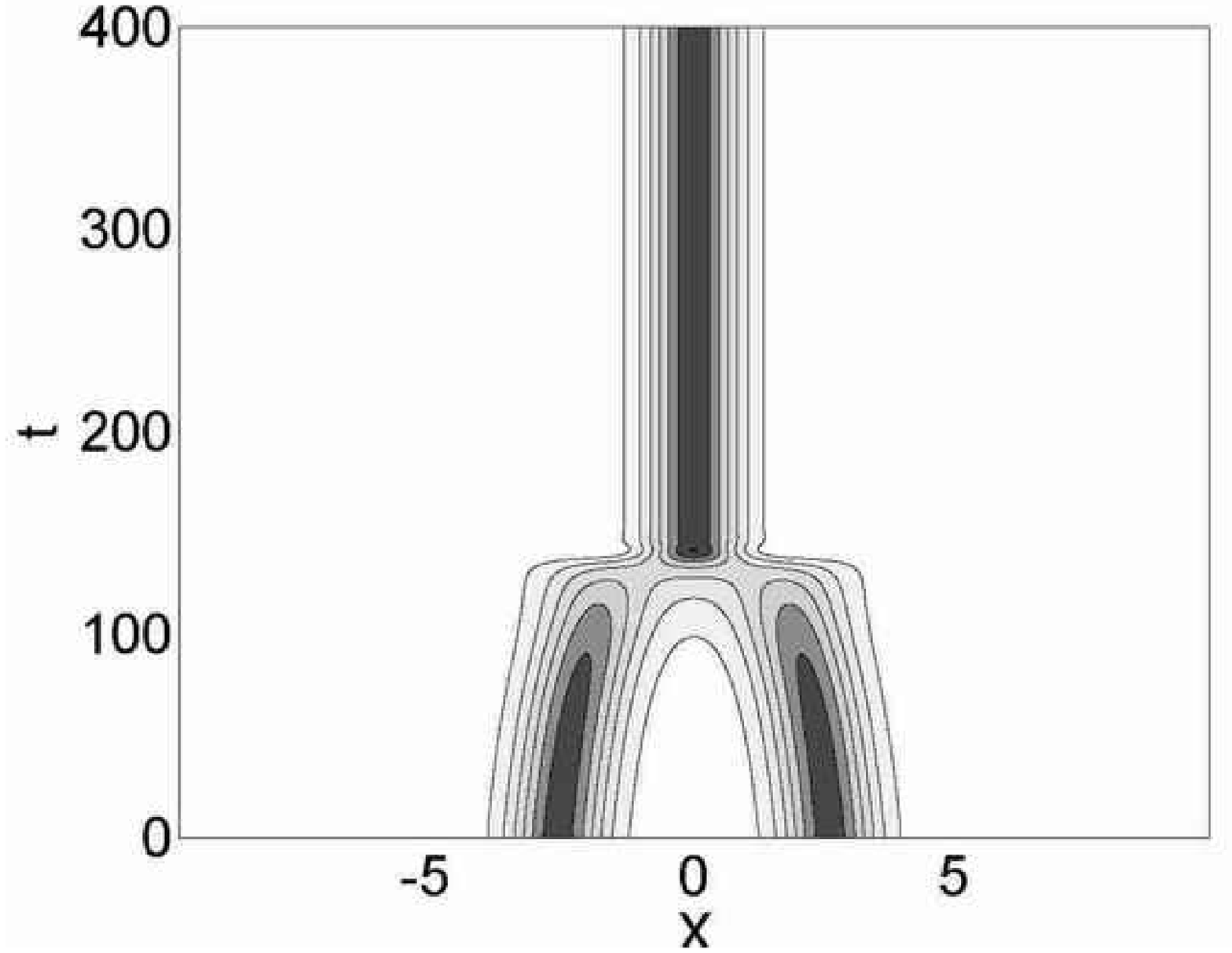} \\[0.4cm]
\mbox{\bf (a)} & \mbox{\bf (b)}\end{array}$\end{center}
\caption{Direct merger of two solitons into a static soliton, in
the FF (a) and SH (b) components. The parameters are $q=0.195$,
$\protect\alpha _{1}=0.2 $, $\protect\alpha _{0}=\left(
\protect\alpha _{0}\right) _{\mathrm{exact}}=\allowbreak 0.297$,
and the initial separation between the solitons is $X_{0}=5.2$}
\label{fig:Double_sol_5}
\end{figure}

\begin{figure}[tbph]
\begin{center}
$\begin{array}{c@{\hspace{0.5in}}c}
\includegraphics[width=2.5in]{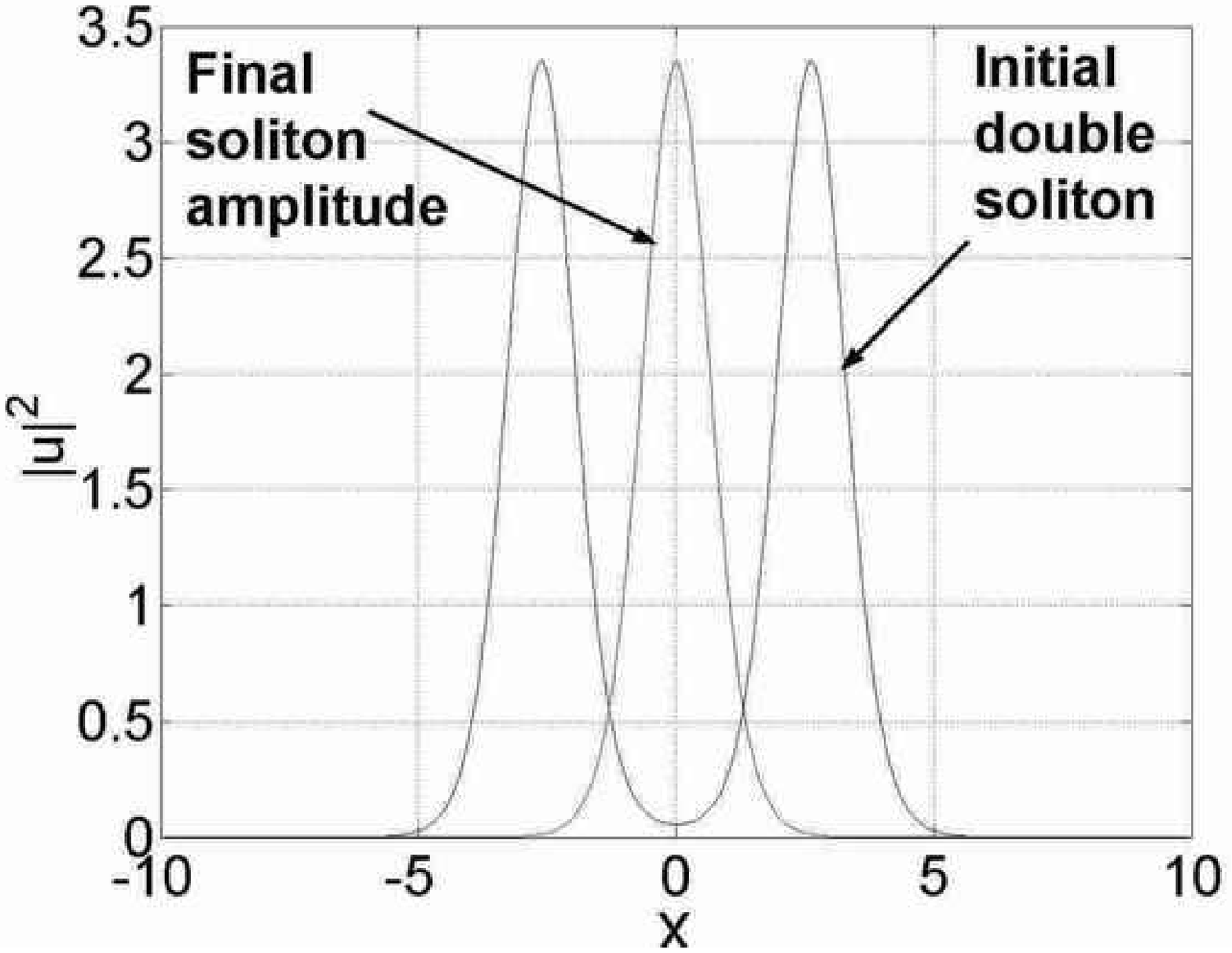} & \includegraphics[width=2.5in]{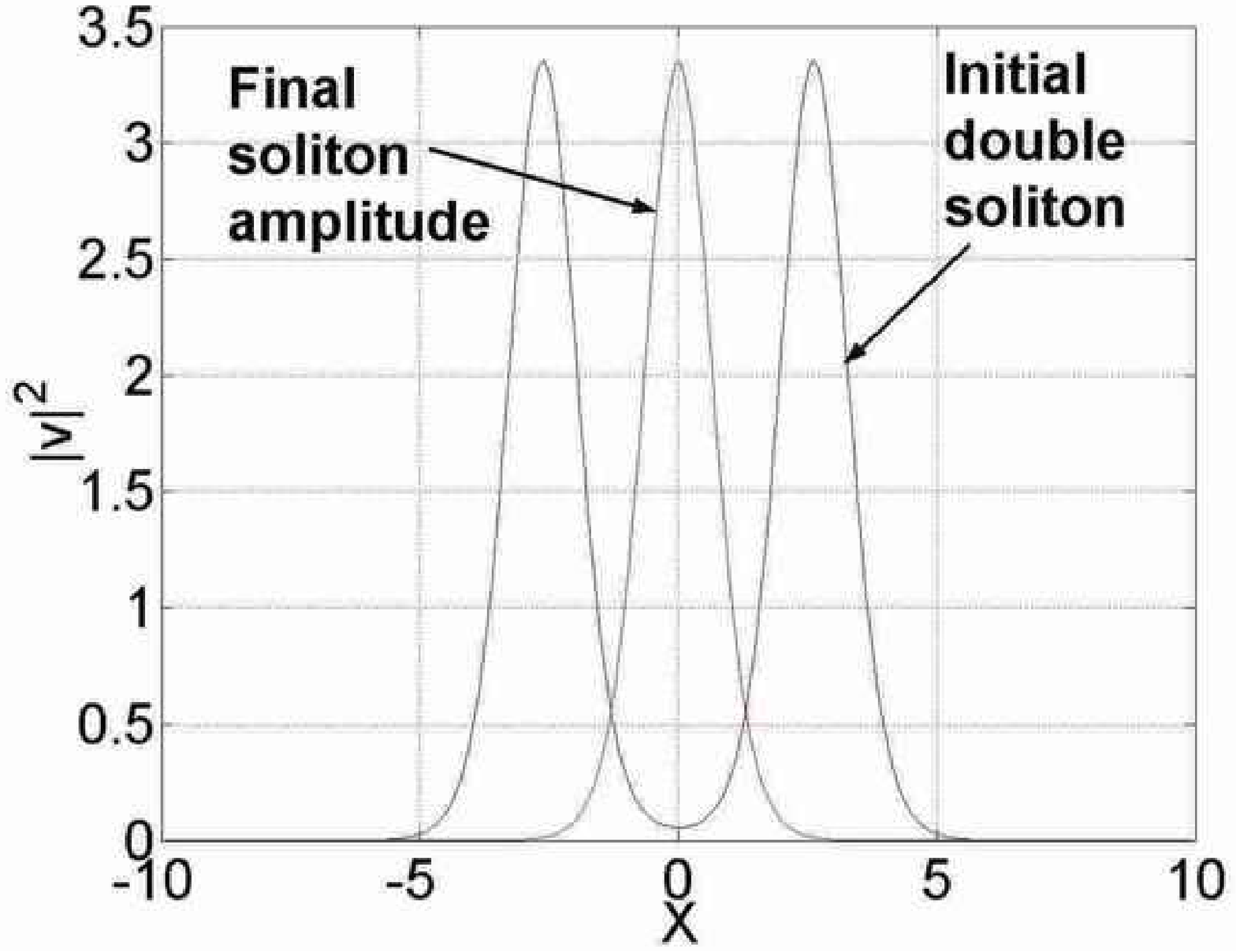} \\[0.4cm]
\mbox{\bf (a)} & \mbox{\bf (b)}\end{array}$\end{center}
\caption{The comparison of the initial and final wave-field
profiles in the FF\ (a) and SH (b) components in the same case
which is shown in Fig. \protect\ref{fig:Double_sol_5}.}
\label{fig:comparison}
\end{figure}

If $\alpha _{1}$ is larger, so that the soliton does not support intrinsic
oscillatory modes, two solitons, even separated by a relatively large
distance, fuse into a single soliton which emerges in the stationary state
(without intrinsic vibrations), as shown in Fig. \ref{fig:Double_sol_5}. The
fact that the final soliton is identical to each initial one is obvious from
Fig. \ref{fig:comparison}, which compares the initial field configuration in
both component and its eventual shape.

\section{Conclusion}

In this work, we have introduced a model of a lossy
second-harmonic-generating ($\chi ^{(2)}$) cavity driven by a pump wave at
the third harmonic, which gives rise to a new type of driving terms,
characterized by the cross-parametric gain. The equation for the
fundamental-frequency wave may also contain a quadratic self-driving term,
which is generated by the $\chi ^{(3)}$ nonlinearity.

Unlike previously studied phase-matched models of $\chi ^{(2)}$ cavities
driven through down- or upconversion, the present model admits the exact
analytical solution for the soliton, at the specially chosen value of the
gain parameter. Two general families of soliton solutions were found in a
numerical form, one of which is a continuation of the exact analytical
solution. At given values of the parameters, one soliton is stable, and one
is not. They swap the stability via a bifurcation which occurs at a critical
value of the mismatch parameter. Full stability regions of the solitons were
identified by means of numerical computation of the corresponding
eigenvalues for small perturbations (stability conditions for the zero
solution, which is a necessary ingredient of the full conditions for the
soliton's stability, were found in an analytical form). The stability of the
solitons was also verified in direct simulations, with the conclusion that
the unstable soliton rearranges into the stable one (which may appear in the
form of a breather), or into a delocalized state, or decays to zero.
Additionally, it was found that steadily moving solitons do not exist in the
present model. If the soliton is initially boosted, it either comes to a
halt, or, if pushed too hard, gets destroyed (possibly, via splitting into
two pulses).

Interactions between initially separated solitons were also investigated by
dint of systematic direct simulations. It was found that stable solitons
always merge into a single one. In the system with weak loss, the final
solitons appears in an excited form (the breather), and then slowly relaxes
to the static configuration. If the loss is stronger, the final soliton
emerges in the stationary form.

The model introduced in this work can be further investigated in various
directions. In particular, a two-dimensional version of this cavity model
may be an interesting subject.


\begin{thebibliography}{99}
\bibitem{ReviewItaly} A. Gatti, A. Lugiato, L. Spinelli, G. Tissoni, M.
Brambilla, P. Di Trapani, F. Pratti, G. L. Oppo, and A. Berzanskis, Chaos,
Solitons and Fractals \textbf{10}, 875 (1999).

\bibitem{WeissReview} 
C. O. Weiss, M. Vaupel, K. Staliunas, G. Slekys, and V. B. Taranenko, Appl.
Phys. B (Lasers and Opt.) \textbf{68}, 151 
(1999); C. O. Weiss, G. Slekys, V. B. Taranenko, K. Staliunas, and R.
Kuszelewicz, 
V. B. Taranenko, G. Slekys, and C. O. Weiss, Chaos \textbf{13}, 
777 
(2003).

\bibitem{Lugiato:downconversion} G. L. Oppo, M. Brambilla, and L. A.
Lugiato, Phys. Rev. A \textbf{49}, 2028 (1994).

\bibitem{Trillo} S. Trillo and M. Haelterman, Opt. Lett. \textbf{23}, 1514
(1998).

\bibitem{Falk:down} C. Etrich, D. Michaelis, and F. Lederer, J. Opt. Soc.
Am. B \textbf{19}, 792 (2002).

\bibitem{Longhi} S. Longhi, Phys. Scripta. \textbf{56}, 611 (1997).

\bibitem{Staliunas} K. Staliunas and V. J. Sanchez-Morcillo, Opt. Commun.
\textbf{139}, 306 (1997).

\bibitem{Skryabin}
D. V. Skryabin, Phys. Rev. E \textbf{60}, R3508 
(1999).

\bibitem{Falk:up} 
D. Michaelis, U. Peschel, C. Etrich, and F. Lederer, IEEE J. Quant. Electr.
\textbf{39}, 255 
(2003).

\bibitem{Alan} 
D. V. Skryabin, A. R. Champneys, and W. J. Firth, Phys. Rev. Lett. \textbf{84}, 463 
(2000).

\bibitem{stability-experiment}
V. B. Taranenko, M. Zander, P. Wobben, and C. O. Weiss, Appl. Phys. B
(Lasers and Opt.) \textbf{69}, 337 
(1999).

\bibitem{moving} D. V. Skryabin and A. R. Champneys, Phys. Rev. E \textbf{63}, 066610 (2001); S. Fedorov, D. Michaelis, U. Peschel, C. Etrich, D. V.
Skryabin, N. Rosanov, and F. Lederer, Phys. Rev. E \textbf{64}, 036610
(2001); 
A. Barsella, C. Lepers, M. Taki, and M. Tlidi, Opt. Commun. \textbf{232},
381 
(2004).

\bibitem{interactions}
D. V. Skryabin and W. J. Firth, Opt. Lett. \textbf{24}, 1056 
(1999).

\bibitem{type-II}
S. Longhi, Opt. Lett. \textbf{23}, 346 
(1998); 
R. A. Fuerst, M. T. G. Canva, G. I. Stegeman, G. Leo, and G. Assanto, Opt.
Quant. Electr. \textbf{30}, 907 
(1998).

\bibitem{3-wave-solitons}
E. Ibragimov, A. A. Struthers, D. J. Kaup, J. D. Khaydarov, and K. D.
Singer, Phys. Rev. E \textbf{59}, 6122 
(1999).

\bibitem{QPM} 
S. C. Rodriguez, J. P. Torres, L. Torner, M. M. Fejer, J. Opt. Soc. Am.
\textbf{19}, 1396 
(2002).

\bibitem{2D}
G. L. Oppo, A. J. Scroggie, and W. J. Firth, J. Opt. B Quant. Semicl. Opt.
\textbf{1}, 133 
(1999).

\bibitem{multi-pixel} 
S. Minardi, A. Varanavicus, A. Piskarskas, and P. Di Trapani, Opt. Commun.
\textbf{224}, 301 
(2003).

\bibitem{LeBerre} M. Le Berre, E. Ressayre, and A. Tallet,
J. Opt. B - Quant. Semicl. Opt. \textbf{2},\textbf{\ }
347
(2000).

\bibitem{multistep}
Y. S. Kivshar, T. J. Alexander, and S. Saltiel, Opt. Lett. \textbf{24}, 759
(1999).

\bibitem{exact-solution-cascading-limit} G. J. Valcarcel, E. Roldan, and K.
Staliunas, Opt. Commun. \textbf{181}, 207
(2000).

\bibitem{Barash} I. V. Barashenkov, M. M. Bogdan, and V. I. Korobov,
Europhys. Lett. \textbf{15}, 113 (1991).

\bibitem{Lucian} L.-C. Crasovan, B. A. Malomed, D. Mihalache, D. Mazilu, and
F. Lederer, Phys. Rev. E \textbf{62}, 1322 (2000).

\bibitem{PhysicaD} B.A. Malomed,
Physica D \textbf{29}, 155 
(1987).

\bibitem{Cai} D. Cai, A. R. Bishop, N. Gr\o nbech-Jensen, and B.A. Malomed,
Phys. Rev. E \textbf{49}, 1677 (1994).
\end{thebibliography}
\end{document}